\documentclass[aps,prb,reprint,amsfonts,amsmath,amssymb]{revtex4-2}
\usepackage[compat=1.1.0]{tikz-feynhand}
\usepackage{color}
\usepackage{dcolumn}% Align table columns on decimal point
\usepackage{bm}% bold math
\usepackage[version=3]{mhchem}
\usepackage{mathtools}
\usepackage{newtxtext}
\usepackage[varg]{newtxmath}
\usepackage[normalem]{ulem}
\usepackage{lipsum}
\usepackage{braket}
\usepackage{comment}
\usepackage[final]{hyperref}
\hypersetup{colorlinks=true,linkcolor=blue,citecolor=blue,urlcolor=blue}
\newcommand{\tr}[1]{\textrm{#1}}

\begin{document}

\preprint{APS/123-QED}

\title{Numerical study of resonant inelastic X-ray scattering at the oxygen $K$-edge in insulating cuprates}
% Force line breaks with \\

%>>>>>>>>>>>>>>>>>>>>>>>>>>>>>>>>>>>>>>>>>>>>>>>>>>>>>>>>>>>>>>>>>>>>>>>
%                                 authors
%>>>>>>>>>>>>>>>>>>>>>>>>>>>>>>>>>>>>>>>>>>>>>>>>>>>>>>>>>>>>>>>>>>>>>>>

\newcommand{\TohokuUniv}{$^2$Department of Physics, Tohoku University, Sendai 980-8578, Japan}
\newcommand{\TohokuUnivApp}{$^1$Department of Applied Physics, Tohoku University, Sendai 980-8577, Japan}

\author{$^1$Yukihiro Matsubayashi and $^2$Sumio Ishihara}
\affiliation{\TohokuUnivApp}
\affiliation{\TohokuUniv}

%<<<<<<<<<<<<<<<<<<<<<<<<<<<<<<<<<<<<<<<<<<<<<<<<<<<<<<<<<<<<<<<<<<<<<<

\date{\today}% It is always \today, today,

\begin{abstract}
We investigate resonant inelastic X-ray scattering (RIXS) at the O $K$-edge in insulating cuprates by means of three methods: cluster perturbation theory (CPT), Hartree--Fock approximation (HFA) and exact diagonalization (ED) method.
We consider the three-band Hubbard model and show the overall momentum-dependence of the Zhang--Rice singlet (ZRS) excitation and charge-transfer excitation by the CPT combining with the perturbation scheme.
A comparison of the RIXS spectra calculated using CPT and HFA reveals different momentum-dependencies through the changes in the properties of the upper Hubbard band and the ZRS band.
These findings are supported by analyses using the ED method on the RIXS spectra and dynamical charge structure factor.

\end{abstract}

\maketitle

%\tableofcontents

%%%%%%%%%%%%%%%%%%%%%%%%%%%%%%%%%%%%%
\section{Introduction }
The development of high-brilliance synchrotron radiation sources has highlighted the importance of using resonant X-ray inelastic scattering (RIXS) to study the physical properties of materials \cite{Kotani_Shin_2001_review,Ament_2011_review,IshiiTohyama_2013_review,Yao_2018_RIXSReviewNature}.
It is expected that analyses of the electronic states of various materials exhibiting interesting physical properties will develop through complementary measurements of dynamical physical properties by using angle resolved photoemission spectroscopy (ARPES) and neutron scattering. 
RIXS research can be broadly classified into two perspectives: one is elucidation of the dynamical physical properties of novel materials  \cite{Suga2005, Marra2013, Marra2016}; the other is the study of the RIXS process itself \cite{Ament2007, Tsutsui2016, Jia2016, Nocera2018}.
The study of fundamental electron scattering processes continues to be an important problem. RIXS is an indirect process with core-hole excitation, in contrast to direct processes such as ARPES and neutron scattering. A useful feature for analysis is that different substances will show the different RIXS spectra at the X-ray absorption edges.

In particular, to clarify the latter perspective, it is very important to study the characteristics of RIXS by using materials such as high-$T_c$ cuprates as reference materials.
For this purpose, we decided to focus on O $K$-edge RIXS in high-$T_c$ cuprates, especially, charge responses such as charge-transfer (CT) excitation and Zhang-Rice singlet (ZRS) excitation.
Here, the ZRS is a spin singlet state formed locally by two holes in the O 2$p_{x,y}$ and Cu 3$d_{x^2-y^2}$ orbitals.
It is considered to be a quasiparticle in cuprate superconductors \cite{Zhang_Rice_1988,Chen_2013_ZRS_XAS,Chen_2013_ZRS_OXAS,Kung_2016_TBHBM,Monney_2016_ZRS}.
The ZRS excitation is expected to be more clearly observable at the O $K$-edge than at other absorption edges owing to the selective excitation of O $2p$ orbitals.
Actually, the ZRS excitation at the Cu $K$-edge RIXS is unclear, because it involves excitations of not only the Cu 3$d_{x^2-y^2}$ orbital but also the other 3d orbitals \cite{Kim_2002_CuKexp, Ellis_2008_LCOexp, Ellis_2011_CuK}. 

In RIXS phenomena, an incident X-ray excites a core-electron to the valence band, and the subsequent relaxation process emits a scattered X-ray.
Accordingly, the momentum-dependence of elementary excitation spectra can be obtained by precisely measuring the changes in the X-ray's energy ($\Omega = \omega_i - \omega_f)$ and momentum ($\bm{Q}=\bm{q}_i - \bm{q}_f$) in a wide range of ($\bm{Q}, \Omega$) space.
The momentum range of X-rays is wide enough to investigate the elementary excitations of solids in the first Brillouin zone (BZ).
RIXS has a noteworthy feature of measuring finite-$\bm{Q}$ excitation spectra in contrast with optical conductivity experiments measuring the excitations at $\bm{Q} \sim 0$.
Also, element-selective experiments can be performed by tuning the incident X-ray's energy to a specific atomic transition.
In O $K$-edge RIXS, a soft X-ray resonantly excites an electron from the O $1s$ orbital to the O $2p_{x,y}$ orbital.
The momentum of the soft X-ray covers about 40$\%$ of the BZ.
The scattering process can be interpreted as insertion of a test charge into the O 2$p$ orbital.
Accordingly, O $K$-edge RIXS causes excitations originating from density fluctuation of O 2$p$ orbitals.
For example, a ZRS and a doublon (a doubly occupied state of the Cu 3$d_{x^2-y^2}$, $d^{10}$) can be simultaneously excited on adjacent plaquettes by an electron hopping from an O 2$p$ orbitals to one of the surrounding Cu 3$d_{x^2-y^2}$ orbitals \cite{Okada_2002_OKRIXSD4h}.
This process is described by $\ket{d^9 ; d^9} \rightarrow \ket{d^9, \mathrm{ZRS} ; d^{10}}$; we call it \textit{ZRS excitation} in this paper.
Previous studies have shown that the O $K$-edge RIXS in insulating cuprates can also be used to study bimagnon and $d$-$d$ excitations besides ZRS excitation and CT excitation \cite{Okada_2001, Okada_2002_OKRIXSD4h, Okada_2002, Okada_2003, Okada_2006, Okada_2007_OKCuL, Bisogni_2012_BimagnonStudyOKL, Bisogni_2012_BimagnonStudyOKLII, Monney2013, Monney_2016_ZRS, Ishii_2017_OKexp, Schlappa_2018_OK_Twospinon, Yamagami_2020_OKexp, Paris_2021_OK_CGO, Shen_2022_OKRIXS_LNO}.

The exact diagonalization (ED) method has been used to study the RIXS spectra of various strongly correlated materials \cite{TsutsuiTohyama_1999_2dCuK, TsutsuiTohyama_2000_1dCuK, Chen_2010_CuK_RIXS_Theory, Kourtis_2012_1dMott_CuK,Paris_2021_OK_CGO, Shen_2022_OKRIXS_LNO} because it can accurately take into account electronic correlations and core-hole potentials in the complicated RIXS processes.
Application of the ED method is, however, restricted to a small number of lattice sites or simple models such as the single-band Hubbard model.
Thus, the momentum dependence of the O $K$-edge RIXS spectra of insulating cuprates, which requires a model with oxygen sites in addition to copper sites, remains unclear.

In this study, we use the diagrammatic perturbative method developed by Nomura and Igarashi \cite{Igarashi_2005_K-edge1, Igarashi_2006_K-edge2, Igarashi_2013_L-edge} to calculate the O $K$-edge RIXS spectra in the three-band Hubbard model \cite{Yin2008_ZRS_DMFT,Weber2010,Kung_2016_TBHBM,Cui2020_dp} in the insulating phase.
Here, the one-particle Green's functions used to evaluate the RIXS diagrams are calculated within the framework of cluster perturbation theory (CPT) \cite{Senechal_2000, Senechal_2002_CPTHubbard}.
%Such a scheme for the RIXS spectra is calld \textit{CPT-based-approach} in this paper.
CPT is a cluster method that can compute one-particle Green's function of multi-orbital systems.
It gives the overall momentum-dependence of the ZRS excitation and the CT excitation.
We also evaluate the RIXS diagrams using Green's functions based on the Hartree--Fock approximation (HFA) instead of  CPT.
%We call this approach to the RIXS spectra the \textit{HF-based-appraoch}.
We discuss the effects of electronic correlations by comparing the results of these two methods.
We also use the ED method to check whether the local correlation is sufficiently incorporated in the calculation based on CPT.

The rest of this paper is organized as follows.
In Sec. 2, we give a model Hamiltonian, that can describes the process of O $K$-edge RIXS, and briefly introduce the three different methods to calculate RIXS spectra.
In Sec. 3A, we show the RIXS spectra based on CPT, and reveal the overall momentum-dependence of the ZRS excitation and the CT excitation.
In Sec. 3B, we analyze a small 2 by 2 cluster by using the ED method and the Kramers--Heisenberg formula.
In Sec. 3C, we show the electronic structures and the RIXS spectra based on the HFA.
We also discuss the effect of electron correlations on the RIXS spectra by comparing results based on CPT and based on the HFA.
Finally, we summarize our results.

%This paper organized as follows
%%%%%%%%%%%%%%%%%%%%%%%%%%%%%%%%%%%%%
\section{Model}
We will use the $dp$ Hamiltonian to examine the RIXS spectra at the O $K$-edge.
The $3d_{x^2-y^2}$ and $2p_{x,y}$ orbitals are taken  account at each Cu and O sites, respectively, in two-dimensional CuO$_2$ plane.  
The Hamiltonian is divided into a one-body part $({\cal H}_{0})$ and a interaction part $({\cal H}_U)$ as 
\begin{align}
{\cal H}_{dp}={\cal H}_{0}+{\cal H}_{U} . 
\label{eq:hpd}
\end{align}
The explicit forms are given by 
\begin{align}
{\cal H}_0&=
  \varepsilon^d \sum_{i  \sigma} d_{i \sigma}^\dagger d_{i \sigma}
+\varepsilon^p \sum_{i \alpha \sigma} p_{i+\alpha \sigma}^\dagger p_{i+\alpha  \sigma} 
\nonumber \\
&+\sum_{i \sigma} t^{dp}
d_{i \sigma}^\dagger \left (-p_{i+x  \sigma} +p_{i+y  \sigma}+p_{i-x \sigma}-p_{i-y \sigma}\right ) + \mathrm{H.c.}  
\nonumber \\
&+\sum_{i \sigma} t^{pp}
\left( p_{i+y \sigma}^\dagger -p_{i-y \sigma}^\dagger \right )
\left (p_{i+x \sigma}- p_{i-x  \sigma} \right ) + \mathrm{H.c.} , 
\label{eq:h0}
\end{align}
and 
\begin{align}
{\cal H}_U=U^d \sum_i n_{i \uparrow}^d n_{i \downarrow}^d
+U^p  \sum_{i \alpha } n_{i+\alpha \uparrow}^p n_{i+\alpha  \downarrow}^p .
\label{eq:hu}
\end{align}
Here, We have introduced the creation (annihilation) operators $d_{i\sigma}^\dagger$ ($d_{i\sigma}$) of the $3d_{x^2-y^2}$ hole with spin $\sigma(=\uparrow, \downarrow)$ at site $i$, and 
those $p^\dagger_{i \pm{\alpha} \sigma}$ ($p_{i\pm\alpha  \sigma}$) of the $2p_\alpha$ hole with orbital $\alpha(=x, y)$ and spin $\sigma$ at  site $i \pm\alpha$. 
The position operators at site $i$ and $i \pm \alpha$ are $\bm{r}_{i}$ and $\bm{r}_i+a \bm{e}_\alpha /2$, respectively, where $\bm{e}_\alpha$ is the unit vector along the $\alpha$ direction, and $a$ is the lattice constant. 
The number operators are defined by $n^d_{i \sigma}=d^\dagger_{i \sigma} d_{i \sigma}$ and $n^p_{i+\alpha \sigma}=p^\dagger_{i+\alpha \sigma}p_{i+\alpha \sigma}$. 
%The transfer integrals, $t_{ij}^{pd}$ and $t_{ij}^{pp}$, 
%are factorized as $t_{ij}^{pd}=t^{pd} \alpha_{ij}$ and $t^{pp}_{ij}=t^{pp}\alpha_{ij}$, respectively, with the sign factor $\alpha_{ij}=\pm 1$, 
At each oxygen site, we consider either of the two O $2p$ orbitals forming the $\sigma$ bonds with the neighboring Cu $3d_{x^2-y^2}$ orbitals. 
In this model, the relation between one-particle energy levels in the electron and hole picture is given by $\epsilon^{d}_{\mathrm{hole}}=-(\epsilon^d + U^d)$, $\epsilon^{p}_{\mathrm{hole}}=-(\epsilon^p + U^p)$.
The charge-transfer energy in the hole-picture in terms of ED and CPT is $\epsilon^{p}_{\mathrm{hole}} - \epsilon^{d}_{\mathrm{hole}}$ = 3~eV, while in terms of the HFA, it is determined self-consistently and we set $\epsilon^d + U^d \braket{n^d}/2 - \epsilon^p =-0.5$~eV, which roughly reproduces the experimentally observed Mott gap energy.

\begin{figure}[t]
\begin{center}
\includegraphics[width=0.6\columnwidth,clip]{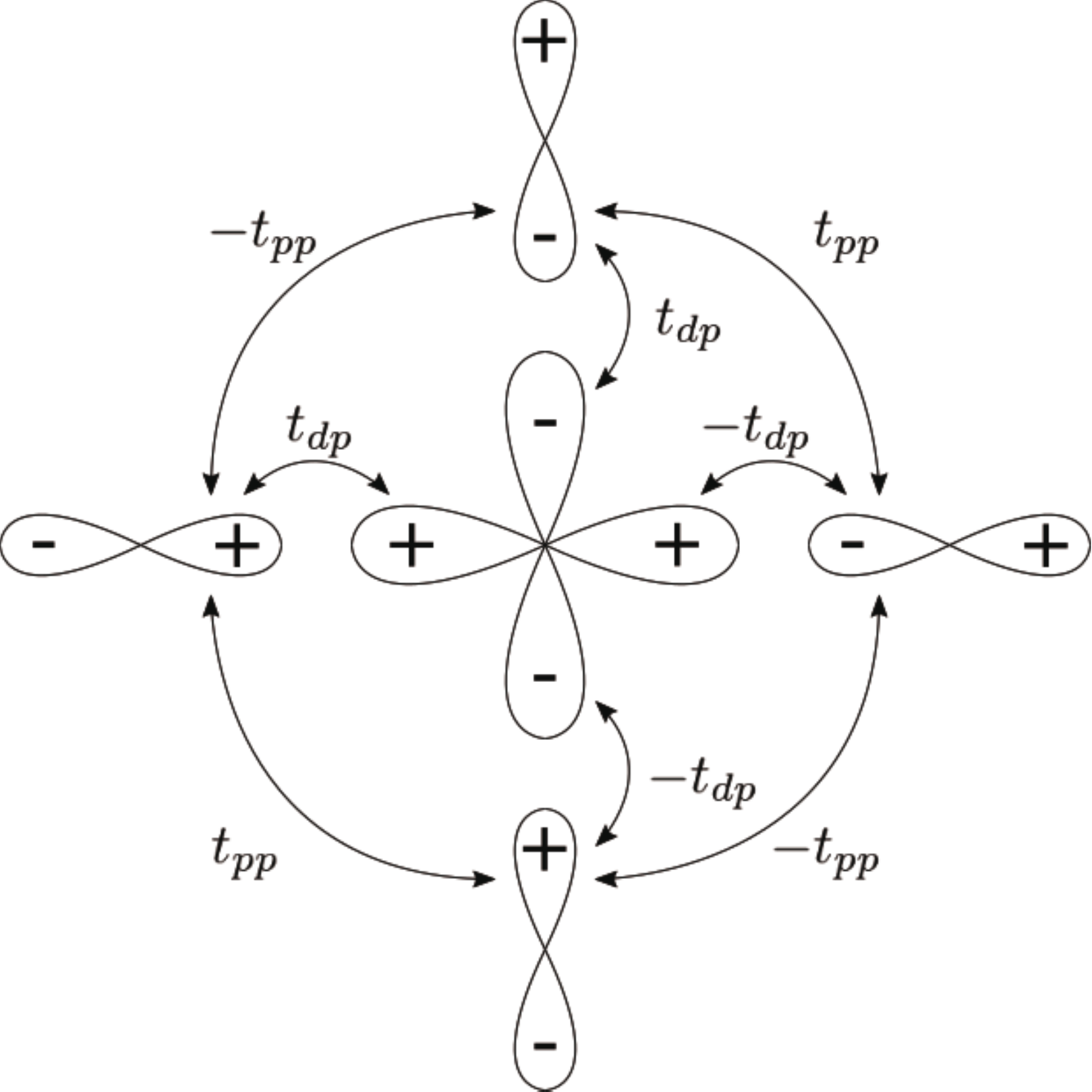}
\end{center}
\caption{
Schematic illustration of the CuO$_2$ plane, and the orbitals and hopping integrals in ${\cal H}_{dp}$
( Eq.~(\ref{eq:hpd})). 
}
\label{fig:cuo2} 
\end{figure}

A schematic view of the orbitals and transfer integrals are shown in Fig.~\ref{fig:cuo2}. 
In the intermediate state of the RIXS processes, an O $1s$ core hole is created, to which the following core-hole potential Hamiltonian is related: 
\begin{align}
{\cal H}_{c}=
\varepsilon^s \sum_{i \alpha \sigma}  s_{i+\alpha \sigma}^\dagger s_{i+ \alpha \sigma} 
+V_{c} \sum_{i \alpha} n^p_{i+\alpha } n^s_{i+\alpha} , 
\end{align}
where $s_{i+\alpha \sigma}^\dagger$ ($s_{i+\alpha \sigma}$) is the creation (annihilation) operator of the O $1s$ hole with spin $s$ at site $i+\alpha$, and 
$n_{i+\alpha}^s=\sum_\sigma s_{i+\alpha \sigma}^\dagger s_{i+\alpha \sigma}$ and 
$n_{i+\alpha}^p=\sum_{\sigma} n_{i+\alpha \sigma}^p$ are the number operators. 
The second term represents the core-hole potential. 

%-------------------------------------------------------------------
%------------------------------------------------------------------

The interaction between electrons and photons causes the O $1s \rightarrow 2p$ transition, which is formulated within the dipole approximation as 
\begin{align}
{\cal H}_{ep}=\sum_{\bm{k} \lambda } 
\sum_{i \alpha \sigma}  w_{\bm{k} \alpha} e^{i \bm{k}\cdot \bm{r}_{i+\alpha}} e_{\bm{k} \lambda}^\alpha
p_{i+\alpha \sigma}^\dagger s_{i+\alpha \sigma}  c_{\bm{k} \lambda} 
+ \mathrm{H.c.} ,
\label{eq:eleph}
\end{align}
where $c^\dagger_{\bm{k}\lambda}$ ($c_{\bm{k} \lambda}$) is the creation (annihilation) operator of a photon with momentum $\bm{k}$ and polarization $\lambda(=1,2)$, $\bm{e}_{\bm{k} \lambda}$ is the polarization vector, and $w_{\bm{k} \alpha}$ is the dipole matrix element. 
For convenience, we introduce the dipole operator: 
\begin{align}
h_{\bm{k} \lambda}=\sum_{i \alpha s} 
w_{\bm{k} \alpha} e^{i \bm{k}\cdot \bm{r}_{i+\alpha}} e_{\bm{k} \lambda}^\alpha
p_{i+\alpha s}^\dagger s_{i+\alpha s}   , 
\end{align} 
and we rewrite ${\cal H}_{ep}$ as
\begin{align}
{\cal H}_{ep}=\sum_{\bm{k} \lambda}  h_{\bm{k} \lambda} c_{\bm{k} \lambda} + \mathrm{H.c.} . 
\end{align}

%%%%%%%%%%%%%%%%%%%%%%%%%%%%%%%%%%%%%
\section{Method}
We will investigate the RIXS spectra by means of the ED method, CPT and the HFA.
The ED method when it is used in combination with the Kramers-Heisenberg formula offers numerically exact results, but has high computational costs.
CPT gives the one-particle Green's functions as a good approximation of those in the thermodynamic limit.
The RIXS diagrams evaluated by CPT Green's function have higher momentum resolution than those of the ED method but the intermediate states in the RIXS process are treated as perturbations.
Finally, in order to reveal how electronic correlations affect the RIXS spectra, the one-particle Green's functions can be calculated within the standard HFA by assuming antiferromagnetic (AF) order.

\subsection{Kramers--Heisenberg formula}
We consider X-ray scattering where the initial, intermediate and final electronic states are given by 
$| 0 \rangle$, $|m \rangle$, and $| f \rangle$, with energies $E_0$, $E_m$, and $E_f$, respectively, 
where $| 0 \rangle $ and $E_0$ are the ground state and its energy, respectively. 
The incident and scattered X-rays characterized by the frequency, momentum and polarization are represented by 
$(\omega_i(=c|\bm{k}_i|), \bm{k}_i, \lambda_i)$ and $(\omega_f(=c |\bm{k}_f|), \bm{k}_f, \lambda_f)$, respectively, where $c$ is the velocity of light. 
Accordingly, the resonant  X-ray scattering intensity is given by the Kramers-Heisenberg formula \cite{Ament_2011_review, IshiiTohyama_2013_review, Okada_2002_OKRIXSD4h}
\begin{align}
I_{\rm RIXS}=\sum_{f}  \left | \sum_m \frac{\langle f| D_f^\dagger | m \rangle \langle m| D_i | 0 \rangle}
{E_0 + \omega_{i} - E_m + i\Gamma}\right |^2 
\delta\left( E_0 + \omega_i - E_f - \omega_f \right) , 
\label{eq:KHF}
\end{align}
where
\begin{align}
D_{i(f)}=\sum_{j \sigma} e^{i \bm{k}_{i (f)} \cdot \bm{r}_j}  
p_{j+\bm{e}_{\lambda_{i(f)} } \sigma}^\dagger s_{j+\bm{e}_{\lambda_{i(f)}} \sigma} , 
%+H.c.  , 
\end{align}
is the dipole operator describing the transition between the O $2p_{\lambda_{i(f)}}$ and $1s$ orbitals, and $\Gamma$ is the core hole damping factor. 
Both $D$ and ${\cal H}_{ep}$ describe the dipole transition,
but ${\cal H}_{ep}$ includes the matrix element of the 1s $\rightarrow$ 2p dipole transition.
%In Eq.~(\ref{eq:KHF}), the resonant term is only considered in the standard Kramers--Heisenberg formula. 
In the same way, the X-ray absorption spectra are given by 
\begin{align}
\label{eq:XAS}
I_{\rm XAS}=-\frac{1}{\pi}{\rm Im} \langle 0 |  
D_i^{\dagger} \frac{1}{E_0 + \omega_i - {\cal H}_{pd} - {\cal H}_{c} + i\Gamma} D_i 
|0 \rangle .
\end{align}

\subsection{Perturbative approach based on Keldysh Green's function}
\label{sec:KGF}
We will follow the formulation given in Ref.~\cite{Igarashi_2005_K-edge1, Igarashi_2006_K-edge2, Igarashi_2013_L-edge} for the O $K$-edge. 
The RIXS intensity can be calculated as the transition probability per unit time from the initial electron and photon states to the final ones:
\begin{align}
W_{\bm{k}_f \lambda_f; \bm{k}_i \lambda_i}=\lim_{t_0 \rightarrow \infty}\frac{d}{dt_0} 
P_{\bm{k}_f \lambda_f; \bm{k}_i \lambda_i} (t_0). 
\label{eq:WWW}
\end{align}
The probability that a photon with momentum $\bm{k}_f$ and polarization $\lambda_f$ is found at time $t_0$ is 
\begin{align}
P_{\bm{k}_f \lambda_f; \bm{k}_i \lambda_i}(t_0)&=
\langle \Phi |U(-\infty, t_0) c_{\bm{k}_f \lambda_f}^\dagger(t_0) 
\nonumber \\
&\times
c_{\bm{k}_f \lambda_f}(t_0) U(t_0, -\infty)| \Phi \rangle , 
\end{align}
where $U(t, t')$ is the time-evolution matrix, and $|\Phi \rangle=c^\dagger_{\bm{k}_i \lambda_i}|0 \rangle \otimes |0 \rangle_{\rm ph}$ with the vacuum state of photon $|0 \rangle _{\rm ph}$. 
By expanding $U$ with respect to the electron-photon interaction ${\cal H}_{ep}$ in Eq.~(\ref{eq:eleph})
up to second order, we have 
\begin{align}
P_{\bm{k}_f \lambda_f; \bm{k}_i \lambda_i}(t_0) &=
\int_{-\infty}^{t_0} du \int_{-\infty}^{u} dt \int_{-\infty}^{t_0} du' \int_{-\infty}^{u'} dt'
\nonumber \\
& \times \langle 0 | h_{\bm{k}_i \lambda_i}(t') h_{\bm{k}_f \lambda_f}^\dagger(u') 
h_{\bm{k}_f \lambda_f}^\dagger(u) h_{\bm{k}_i \lambda_i}(t) |0 \rangle
\nonumber \\ 
&\times e^{i \omega_i(t'-t)} e^{-i \omega_f(u'-u)} ,
\label{eq:PPP}
\end{align}
where $h_{\bm{k} \lambda}(t)$ is the Heisenberg representation of $h_{\bm{k} \lambda}$.

%------------------------------------------------------------------------------
\begin{figure}[t]
\begin{tikzpicture}[scale = 0.7] 
\begin{feynhand}
    \vertex [photon] (in)  at (-3.5,3);
    \vertex [photon] (out) at (3.5,3);
    \vertex [particle] (v1) [label=above:$t$] at (-1.5, 1.5);
    \vertex [particle] (v2) [label=above:$0$] at (1.5, 1.5);
    \vertex [particle] (v3) [label=below:$u'$] at (1.5, -1.5);
    \vertex [particle] (v4) [label=below:$t'$] at (-1.5, -1.5);
    \vertex [photon] (rin) at (3.5,-3);
    \vertex [photon] (rout) at (-3.5,-3);
    \propag [chabos] (in) to  [edge label=$\bm{q}_\tr{i} \, \alpha_\tr{i}$] (v1);
    \propag [chabos] (v2) to  [edge label=$\bm{q}_\tr{f} \, \alpha_\tr{f}$] (out);
    \propag [chabos] (rin) to [edge label=$\bm{q}_\tr{f} \, \alpha_\tr{f}$] (v3);
    \propag [chabos] (v4) to  [edge label=$\bm{q}_\tr{i} \, \alpha_\tr{i}$] (rout);
    \propag [anti fermion] (v1) to [edge label=$1s$] (v2);
    \propag [anti fermion] (v2) to [edge label=$2p$] (v3);
    \propag [anti fermion] (v3) to [edge label=$1s$] (v4);
    \propag [anti fermion] (v4) to [edge label=$2p$] (v1);
\end{feynhand}
\end{tikzpicture}
\caption{
Diagram for $W_a$ given in Eq.~(\ref{eq:wa}). 
Solid lines represent the Green's function for the $2p$ and $1s$ electrons and wavy lines represent those of photons.
}
\label{fig:wa} 
\end{figure}
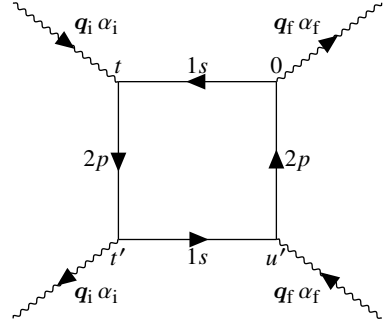
%------------------------------------------------------------------------------

Equation (\ref{eq:WWW}) with Eq.~(\ref{eq:PPP}) is evaluated by the diagrammatic expansion in the Keldysh Green's function formalism. 
Here, we consider four diagrams denoted by $W_a, W_b, W_c, W_{\mathrm{indirect}}$, in the HFA for Cu $K$-edge and Cu $L$-edge in Ref.~\cite{Igarashi_2006_K-edge2}. 
%Since we have checked that the contribution from $W_a$ is much larger than those from the other three diagrams, we only give the explicit form for $W_a$ in Fig.~\ref{fig:wa}:
\begin{align}
W_a&=|w_{\bm{k}_f \lambda_f} w_{\bm{k}_i \lambda_i }|^2 
\int_{-\infty}^{u'} dt' \int_{-\infty}^{\infty} du' \int^{0}_{-\infty} dt 
\nonumber \\
&\times e^{-i(\omega_i+\varepsilon_s+i\Gamma)t'} e^{-i(\omega_f+\varepsilon_s+i\Gamma)u'} e^{-i(\omega_i-\varepsilon_s-i\Gamma)t} 
\nonumber \\
&\times
\frac{2}{N} \sum_{\bm{k} \sigma} \sum_{\alpha \alpha'}  e_{\lambda_i}^{\alpha}  e_{\lambda_f}^{\alpha}  e_{\lambda_i}^{\alpha'}  e_{\lambda_f}^{\alpha'} 
\nonumber \\
&\times
G^{2p+-}_{\alpha \alpha'; \sigma \sigma}(\bm{k+Q}, t', t) G^{2p-+}_{\alpha \alpha'; \sigma \sigma}(\bm{k}, 0, u') ,  \nonumber \\
&
=|w_{\bm{k}_f \lambda_f} w_{\bm{k}_i \lambda_i }|^2
\frac{1}{N} \sum_{\bm{k} \sigma} \sum_{\alpha \alpha'}  e_{\lambda_i}^{\alpha}  e_{\lambda_f}^{\alpha}  e_{\lambda_i}^{\alpha'}  e_{\lambda_f}^{\alpha'}
\nonumber  \\
&
\times
\int^\infty_{-\infty}\dfrac{d\omega}{2\pi}
\left|R(\omega_i, \omega + \Omega)\right|^2
\nonumber  \\
&
\times
G^{2p+-}_{\alpha'\sigma ; \alpha\sigma}(\bm{k}+\bm{Q}, \omega + \Omega)
G^{2p-+}_{\alpha\sigma ; \alpha'\sigma}(\bm{k},\omega) ,
\label{eq:wa}
\end{align}
with
\begin{equation}
\label{eq:Rxy}
    R(x,y) = 1/(x -y + \epsilon_s + i\Gamma)
\end{equation}
where $\Omega=\omega_i-\omega_f$ is transferred energy $\bm{Q}=\bm{q}_i - \bm{q}_f$, and $N$ is the number of unit cells.  
In the Keldysh Green's functions  
$G^{2p \gamma \gamma'}_{\alpha'\sigma' ; \alpha\sigma}(\bm{k}, t', t)$, the superscripts $\gamma$ and $\gamma'$ take $+$ and $-$, which represent the backward and outward time legs, respectively. 
The expressions and the diagrams for $W_b$, $W_c$, and $W_{\mathrm{indirect}}$ are given in Appendix~\ref{sec:diagram}. 
%In the calculations in the HF approximations, we  adopt $W_a$, $W_b$, $W_c$ and $W_{indirect}$, 
%and confirm that $W_a$ is the dominant term. 
%In the CPT introduced in Sec.~\ref{sec:CPT}, we adopt $W_a$ and $W_b$, and the former is the dominant term.

\subsection{Cluster Perturbation Theory}
\label{sec:CPT}
Now let us evaluate the Green's functions in the diagrams by using CPT. 
CPT gives the one-particle Green's functions of multi-orbital systems at low numerical cost and simple procedures.
The Green's functions of a large cluster are constructed from those of a small cluster by incorporating intercluster-hoppings as a perturbation.
Local electron correlation effects are included by using the ED method to solve Green's functions in a small cluster.
In the three-band Hubbard model, the orbital-resolved Green's functions are defined as
\begin{align}
G^{d}_{\mathrm{CPT}}(\bm{k}, \omega)
&=\dfrac{1}{N}\sum_{i,j}e^{-i \bm{k} \cdot(\bm{r}_i - \bm{r}_j)} G_{i, j}(\bm{k}, \omega),\\
G^{p_\alpha}_{\mathrm{CPT}}(\bm{k}, \omega)
&=\dfrac{1}{N}\sum_{i,j}e^{-i \bm{k} \cdot(\bm{r}_{i+\alpha} - \bm{r}_{j+\alpha})} G_{i+\alpha,
j+\alpha}(\bm{k}, \omega)
\end{align}
where $G_{i,j}(\bm{k}, \omega) = [\hat G(\omega)^{-1} - \hat  V(\bm{k})]^{-1}_{i,j}$.
$\hat V(\bm{k})$ is Fourier-transformed intercluster-hoppings.
$\hat G(\omega)$ denotes the numerically exact Green's functions within the small cluster.
Since we have $N$ unit cells which contain the three orbitals, $\hat G(\omega)$ is a $3N \times 3N$ matrix.

\subsection{Hartree--Fock Approximation}
\label{sec:HF}
To describe the AF order within the HFA, the unit cell is defined as \ce{Cu2O4}. 
The HF decompositions are included in the interaction term ${\cal H}_U$ in Eq.~(\ref{eq:hu}) as 
\begin{align}
n_{i \uparrow}^d n_{i \downarrow}^d & \rightarrow 
   \langle n_{i \uparrow}^d \rangle n_{i \downarrow}^d 
+ n_{i \uparrow}^d \langle n_{i \downarrow}^d\rangle
-\langle n_{i \uparrow}^d\rangle \langle  n_{i \downarrow}^d \rangle \nonumber \\
&-   \langle d^\dagger_{i \uparrow} d_{i \downarrow} \rangle d^\dagger_{i \downarrow} d_{i \uparrow} 
-   d^\dagger_{i \uparrow} d_{i \downarrow} \langle d^\dagger_{i \downarrow} d_{i \uparrow} \rangle 
+  \langle d^\dagger_{i \uparrow} d_{i \downarrow} \rangle \langle d^\dagger_{i \downarrow} d_{i \uparrow} \rangle .
\end{align}
Similarly, the HF decomposition is performed for the term $n_{i \uparrow}^p n_{i \downarrow}^p$. 
By introducing the Fourier transform of the fermionic operators defined by  
\begin{align}
d_{\bm{k} s}&=\sqrt{\frac{2}{N}}  \sum_{i} d_{i s} e^{i \bm{r}_i \cdot \bm{k}} , 
\\
p_{\bm{k} \alpha s}&=\sqrt{\frac{2}{N}} 
\sum_{i} p_{i + \alpha s}  e^{i \bm{r}_{i + \alpha} \cdot \bm{k}} , 
\end{align}
the HF Hamiltonian is given by 
\begin{align}
{\cal H}_{HF}=\sum_{\bm{k}} \hat \psi^\dagger_{\bm{k}} \hat h_{\bm{k}} \hat \psi_{\bm{k}} , 
\end{align}
where the bases set is taken to be 
\begin{align}
\hat \psi_{\bm{k}} =\, ^t\left (d_{\bm{k} \uparrow}, p_{\bm{k} x \uparrow}  , p_{\bm{k} y \uparrow}, 
d_{\bm{k} \downarrow}, p_{\bm{k} x \downarrow}  , p_{\bm{k} y \downarrow} \right ) . 
\label{eq:basis}
\end{align}
This Hamiltonian can be diagonalized with a unitary transformation: 
\begin{align}
{\cal H}_{HF}=\sum_{\bm{k}} 
\hat \phi_{\bm{k}^\dagger} \hat \varepsilon_{\bm{k}} \hat \phi_{\bm{k}} ,
\end{align}
where $\hat \phi_{\bm{k}}=\hat U_{\bm{k}}^{-1} \hat \psi_{\bm{k}}$ and 
$\hat \varepsilon_{\bm{k}}(=\hat U_{\bm{k}}^{-1} \hat h_{\bm{k}} \hat U_{\bm{k}})$ is a diagonal matrix. 
Then, the Green's function represented by the basis set in Eq.~(\ref{eq:basis}) is given by 
\begin{align}
\hat G(\bm{k}, \omega)=\hat U_{\bm{k}} 
\left [\frac{1}{\omega-\varepsilon_{\bm{k}} + i \delta\,\mathrm{sgn}(\varepsilon_{\bm{k}})} \right ] \hat U^{-1}_{\bm{k}},
\end{align}
where $\delta$ is a convergent factor.

\section{Results}
\subsection{Cluster perturbation theory}
\begin{figure}[tb]
    \centering
    \includegraphics[width=1\linewidth]{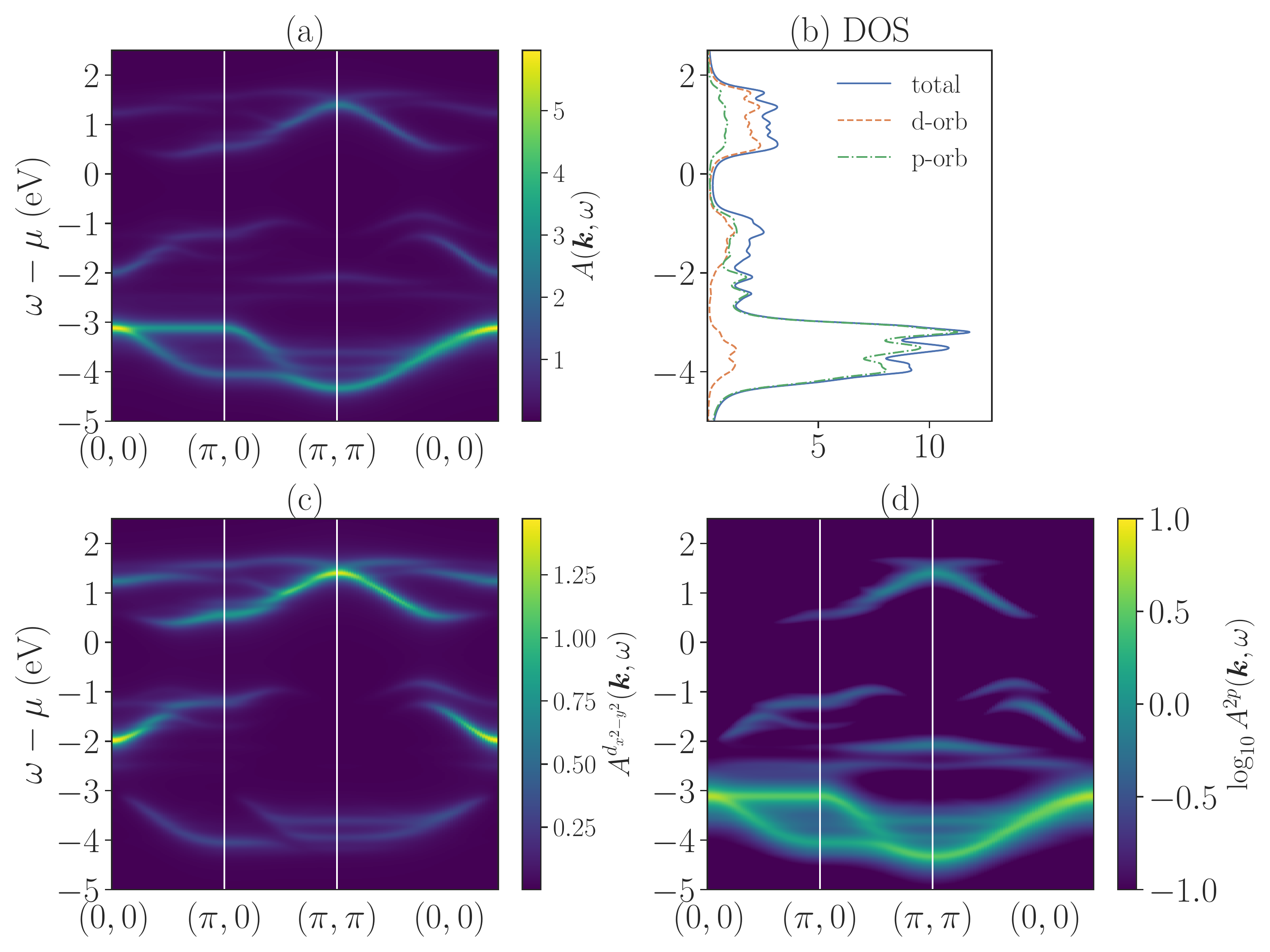}
    \caption{ 
    One-particle Green's functions calculated by using CPT.
    (a) Momentum-resolved spectral function.
    (b) Density of states.
    Momentum- and orbital-resolved spectral functions for (c) Cu $d_{x^2-y^2}$; for (d) O $2p_x, 2p_y$. 
    }
    \label{fig:CPT DOS}
\end{figure}
\begin{figure}[tb]
    \centering
    \includegraphics[width=1\linewidth]{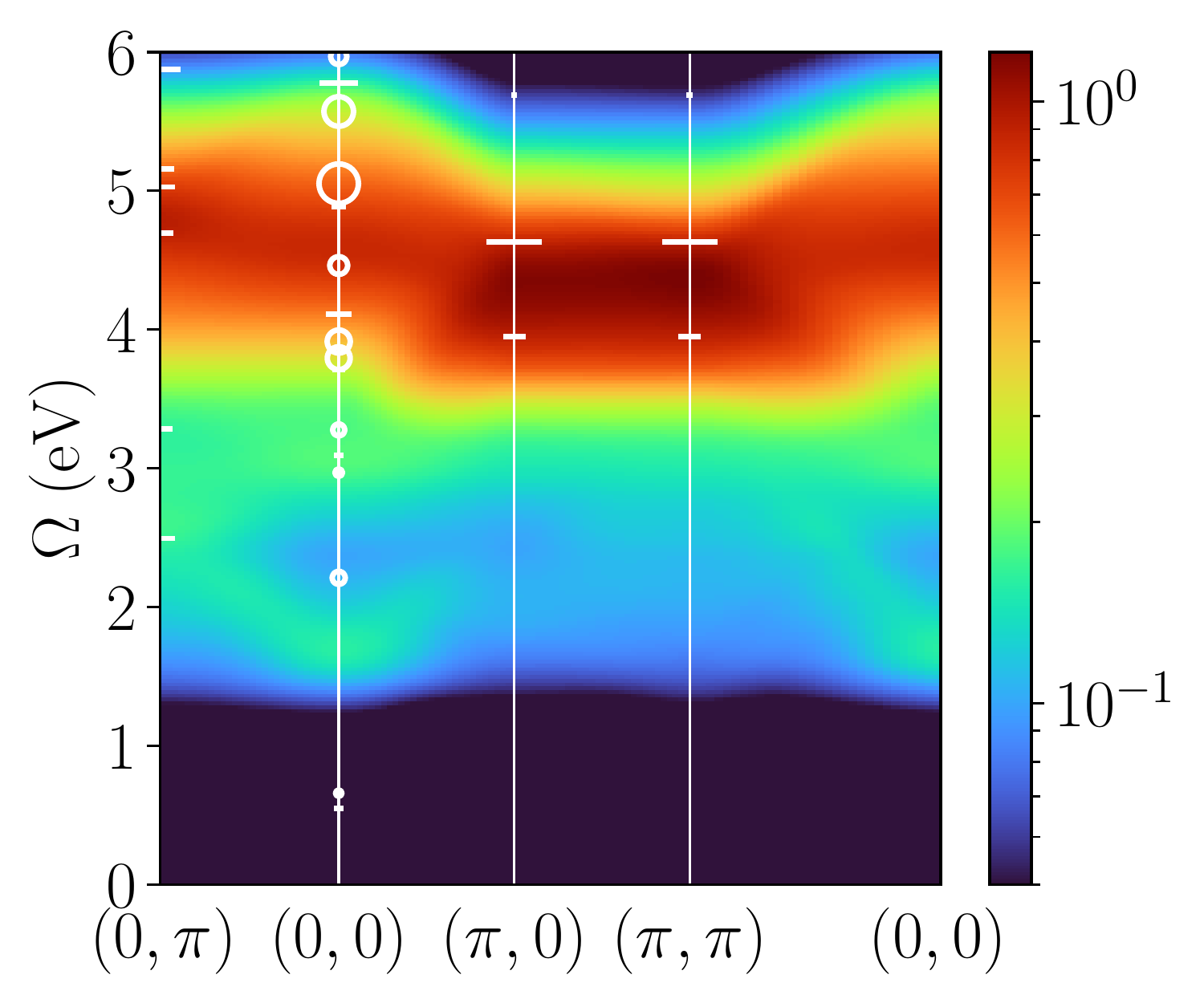}
    \caption{ 
    Oxygen $K$-edge RIXS spectra based on the CPT scheme plotted as a function of the energy loss $\Omega$~(eV) for high-symmetry cuts in the first BZ.
    %(b) Magnified (a) from 0~eV to 3~eV to highlight the ZRS excitation.
    Each of the white horizontal bars and white circles are peak positions extracted from the ED results for $2 \times 2$ and $3 \times 2$ clusters, respectively (see Fig.~\ref{fig:ED XAS DSF RIXS}(c), $S_z=0$.)
    Their marker size is proportional to the intensities of the RIXS spectra.
    }
    \label{fig:CPT RIXS}
\end{figure}
In this subsection, we show numerical results of RIXS spectra using CPT Green's function.
First, we take a look at the one-particle spectra from the imaginary part of the Green's function.
The O $K$-edge RIXS spectra reflect the density of states of the oxygen component, and its peaks arise mainly from the particle-hole excitation.
Thus, we can determine the origin of the RIXS spectra by comparing the RIXS spectra with the one-particle spectral function.

Let us take a 2$\times$2 (Cu$_4$O$_8$) cluster as a reference system of CPT.
We have $t^{dp}=1$~eV, $t^{pp}=0.3$~eV, $\epsilon^{p}_{\mathrm{hole}} - \epsilon^{d}_{\mathrm{hole}}=3$~eV, $U^d=8$~eV and $U^p=4$~eV as a parameter set for typical copper oxides.
The core-hole potential is set as $V_c=5$~eV.
Figures~\ref{fig:CPT DOS}(a) and (b), respectively, show one-particle spectral functions and density of states (DOS) computed using CPT, which include contributions from both the Cu $3d_{x^2-y^2}$ orbital and the O $2p_{x,y}$ orbitals.
The spectral weight projected onto the Cu 3$d_{x^2-y^2}$ orbital and the O $2p_{x,y}$ orbitals are shown in Fig.~\ref{fig:CPT DOS}(c) and (d).
The large energy gap around the Fermi level is due to the strong electronic correlation on the Cu sites.
The upper Hubbard band (UHB) and ZRS band are located above and below the Mott gap, respectively. 
Here, the terminology, ZRS band, corresponds to the one-particle spectra in the energy ranging from $-2$ eV to $-1$ eV in Fig.~\ref{fig:CPT DOS}.
Since the ZRS band has its weight in both Fig.~\ref{fig:CPT DOS}(c) and (d), the band represents the hybridization between the Cu 3$d_{x^2-y^2}$ orbital and the O 2$p_{x,y}$ orbitals.
Although the largest spectral weight in the oxygen component is around $-4$ eV, the DOS in Fig.~\ref{fig:CPT DOS}(b) suggests that the spectral weights for the UHB and ZRS band are large enough to observe the ZRS excitation.

RIXS spectra at the O $K$-edge with the $x$ polarization are shown in Fig.~\ref{fig:CPT RIXS}.
The origin of $\omega_i$ is taken to be at $\epsilon^s$, and $\omega_i$ is chosen as the peak of the DOS of the UHB at 0.6~eV.
%The magnified image of Fig.~\ref{fig:CPT RIXS}(a) for $\Omega \leq 3.5$~eV is shown in Fig.~\ref{fig:CPT RIXS}(b); the momentum dependence of the ZRS excitation is clearly visible around 1.5--2.5~eV.
The momentum dependence of the ZRS excitation is clearly visible around 1.5--2.5~eV.
The ZRS has a maximum intensity at the $\bm{Q}=(0, 0)$ and its energy increases from $(0, 0)$ to the edge of the first BZ.
In this calculation, the ZRS excitation is a particle-hole excitation from the ZRS band to the UHB, which also corresponds to the formation of the doublon and ZRS as mentioned in Sec. I.
The CT excitation around 4~eV has a high intensity.
There is a prominent flat structure along $(\pi, 0)$--$(\pi, \pi)$, which is not observable in experiments because of the limited momentum of soft X-rays.

The momentum dependences of the RIXS spectra along $(0, 0)$--$(\pi, 0)$ and $(0, 0)$--$(0,\pi)$ are different.
This is due to the anisotropy caused by the excitation and decay of the core-hole between the 1$s$ and 2$p_x$ orbitals at the same oxygen site.
Note that the bimagnon excitation observed in experiments is not included in the four diagrams $W_{a,b,c,\mathrm{indirect}}$ considered in this calculation, because the bimagnon excitation is caused by a higher-order process.
However, the energy scale of the bimagnon excitation is much lower than and seprate from those of the ZRS excitation and the CT excitation.

\subsection{Exact diagonalization}
\begin{figure}[tb]
\centering
\includegraphics[width=1.0\columnwidth]{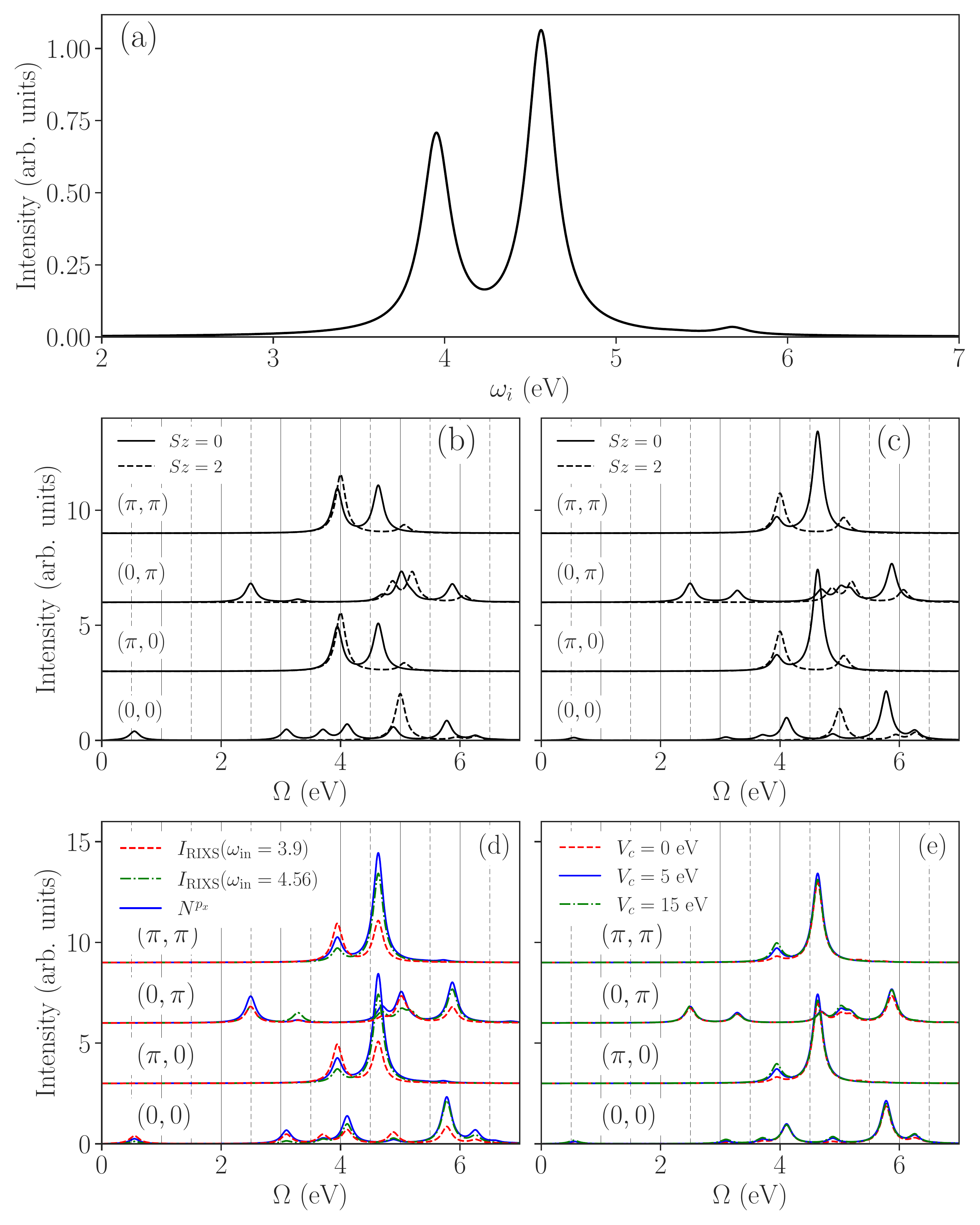}
\caption{ 
    (a) Oxygen $K$-edge XAS spectra in $x$-polarization for the \ce{Cu4O8} cluster.
    There are two resonant peaks at $\omega_i =$ 3.9~eV, 4.56~eV.
    Oxygen $K$-edge RIXS spectra at the absorption edge for (b) $\omega_i=$ 3.9~eV, (c) $\omega_i=$ 4.56~eV.
    (d) O $K$-edge RIXS spectra and dynamical charge structure factor for the O $2p_x$ orbital.
    (e) O $K$-edge RIXS spectra for various core-hole potentials.
}
\label{fig:ED XAS DSF RIXS} 
\end{figure}
\begin{figure*}[t]
    \centering
    \includegraphics[width=1.0\linewidth]{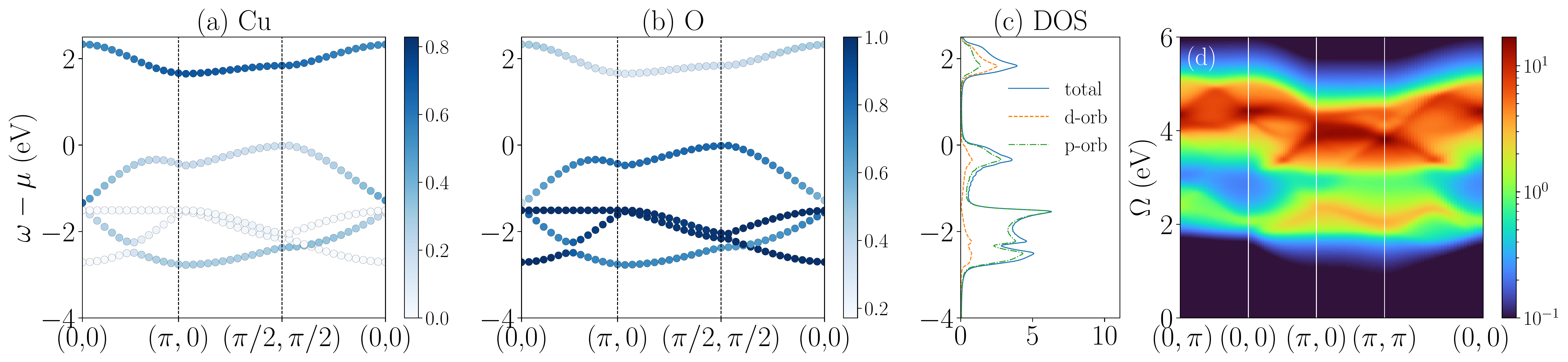}
    \caption{ 
    Orbital-resolved electronic band structure by using the HFA for (a) Cu 3$d_{x^2-y^2}$ and (b) O 2$p$.
    (c) Density of states.
    (d) Calculated RIXS intensity based on the HFA. 
    }
    \label{fig:HF DOS RIXS}
\end{figure*}
To verify the perturbative method used in the previous section, we performed the ED method on 2 $\times$ 2 (Cu$_4$O$_8$) clusters with periodic boundary conditions.
The lifetime for the O 1s core-hole in the intermediate state was taken to be $\Gamma=0.5$~eV.
The Lorentzian broadening $\eta$ for the XAS spectrum was set to 0.1~eV.
The calculation of the RIXS spectra was implemented in two steps:
first, we calculated the XAS spectrum from Eq.~\eqref{eq:XAS} by using the continued fraction expansion method, and determined the resonance incident X-ray's energy from the peak positions of $I_{\mathrm{XAS}}$.
Then, we calculated the RIXS spectra from the Kramers-Heisenberg formula Eq.~\eqref{eq:KHF} by using the bi-conjugate gradient stabilized method (BiCGSTAB).
The parameters were the same as those in CPT.

Figure~\ref{fig:ED XAS DSF RIXS}(a) shows the O $K$-edge XAS spectrum.
The XAS spectrum is composed of two peak structures at $\omega_i \simeq3.9$~eV and 4.56~eV. 
These peak energies correspond to the excitation energies of the core-hole electron at the oxygen site resonantly excited to the O 2$p_x$ component of the UHB.
Figures~\ref{fig:ED XAS DSF RIXS}(b) and \ref{fig:ED XAS DSF RIXS}(c) are the RIXS spectra with the energy of the incident X-ray tuned to the XAS peaks with $\omega_i=$ 3.9~eV and 4.56~eV, respectively.
Note that the components of the elastic scattering have been removed from the spectra.

Next, to assign the ZRS excitation and magnetic components in the RIXS spectra, we calculated the ground states of the $2\times 2$ cluster in restricted Hilbert spaces of which the total spin is fixed to 0 or 2.
In the case of $S_z=2$, the ZRS excitation is forbidden, because all the Cu spins are initially in parallel.
Magnetic excitations such as the bimagnon are also forbidden because only the excitations by an even- or zero-time-spin-flip are allowed in O $K$-edge RIXS for the initial state.
Thus, the peaks around 0.5~eV and the peaks around 3--4~eV indicated by a solid line ($S_z=0$) and momentum $\bm{Q}=(0,0)$ in Fig.~\ref{fig:ED XAS DSF RIXS}(b) are assigned to the bimagnon excitation and the ZRS excitation, respectively.
This is also evidence that the ZRS has a singlet character.
The reason that the bimagnon excitation appears only at $\bm{Q}=(0, 0)$ is probably due to the small cluster size.

Comparing between the solid lines ($S_z=0$) in Figs~\ref{fig:ED XAS DSF RIXS}(b) and \ref{fig:ED XAS DSF RIXS}(c) reveals that the difference in the absorption edges appears in the spectral weights of the ZRS excitation or the CT excitation:
the spectral weight of the ZRS excitation is highest for $\omega_i=3.9$~eV, the spectral weight of the CT excitation is highest for $\omega_i=4.56$~eV.
In Fig.~\ref{fig:CPT RIXS}, the positions of the peaks in Fig.~\ref{fig:ED XAS DSF RIXS}(c) are plotted with white bars while those of the RIXS spectra for the 3$\times$2 cluster are plotted with white circles.
The CT excitations obtained by the ED at $(\pi,0)$ and $(\pi, \pi)$ capture the features of the flat momentum-dependence obtained by CPT in Fig.~\ref{fig:CPT RIXS}.
This indicates that the dominant contributions are properly included in the diagrams treated in the perturbative approach in the previous section, and that the CT excitation at $(\pi,0)$ and $(\pi, \pi)$ can be described even in the four points of the first BZ of the \ce{Cu4O8} ($2\times 2$) cluster.
From the latter point, the CT excitation at $(\pi,0)$ can be assigned to the particle-hole excitation from the O $2p$ band at $(0, \pi)$ to the UHB at $(\pi, \pi)$.
Likewise, the CT excitation at $(\pi, \pi)$ corresponds to the particle-hole excitation from the O $2p$ band at $(0, 0)$ to the UHB at $(\pi, \pi)$.
The ZRS excitation between 2~eV and 3.5~eV also shows good agreement with the CPT results.

In the following, we check whether O $K$-edge RIXS can be interpreted as the dynamical structure factor of O 2$p$ electrons, denoted by $N^{p_x}(\bm{Q},\Omega)$.
In Fig.~\ref{fig:ED XAS DSF RIXS}(d), the RIXS spectra are directly compared with $N^{p_x}$.
The O $K$-edge RIXS spectra are roughly consistent with $N^{p_x}$.
In particular, there is only a slight difference even in the spectral intensities for $\omega_i=4.56$~eV.
In addition, as illustrated in Fig.~\ref{fig:ED XAS DSF RIXS}(e), the RIXS spectra shows the weak dependence on the core-hole potential $V_c$.
A strong dependence would indicates the existence of a complicated scattering process, but in the figure the core-hole potential causes only a slight increase in intensity because the number of holes at O $2_p$ sites is small, as pointed out in Ref~\cite{Okada_2007_OKCuL}. 
These results indicate that the O $K$-edge RIXS spectra can be interpreted basically as the dynamical charge structure factor for the O 2$p$ orbital.

\subsection{Hartree-Fock approximation}
Now let us calculate the one-particle Green's functions and RIXS spectra on the basis of the HFA.
We will use the electron picture rather than the hole picture and the parameter values $t^{dp}=1.0$~eV, $t^{pp}=0.3$~eV, $U^d=8$~eV, $V_c=5$~eV, $\Gamma=0.5$~eV.
The charge-transfer energy can be self-consistently determined as $\epsilon^d + U^d \braket{n^d}/2 - \epsilon^p =-0.5$~eV to reproduce the Mott gap.
Within the HFA, the ground state has AF order, for which the unit cell contains two Cu $3d_{x^2-y^2}$ orbitals and four O $2p$ orbitals.
The electronic band structure and the DOS within the HFA are shown in Figs.~\ref{fig:HF DOS RIXS}(a)--\ref{fig:HF DOS RIXS}(c).
The UHB is above and the ZRS band is below the Fermi energy, and the oxygen band is located around $-4$~eV.
The general shape of the electronic structure is similar to that obtained by CPT.
However, the ZRS band is mainly composed of the O $2p$ component, unlike the CPT case, where the contribution from Cu $3d_{x^2-y^2}$ orbital is comparable with that from O $2p$ orbitals.

The RIXS spectra obtained by the HFA are shown in Figs.~\ref{fig:HF DOS RIXS}(d).
$\omega_i$ was chosen as the peak of the UHB.
The ZRS excitation is around 2~eV, and the CT excitation is around 4~eV.
The ZRS excitation has a maximum intensity at $(\pi, \pi)$ unlike the results based on CPT, and the CT excitation is characterized by several sharp peaks.
Its momentum-dependence, such as $(\pi,0)$--$(\pi,\pi)$, is similar to but slightly different from those of the CPT and ED results.

\begin{figure}[t]
    \centering
    \includegraphics[width=1.0\linewidth]{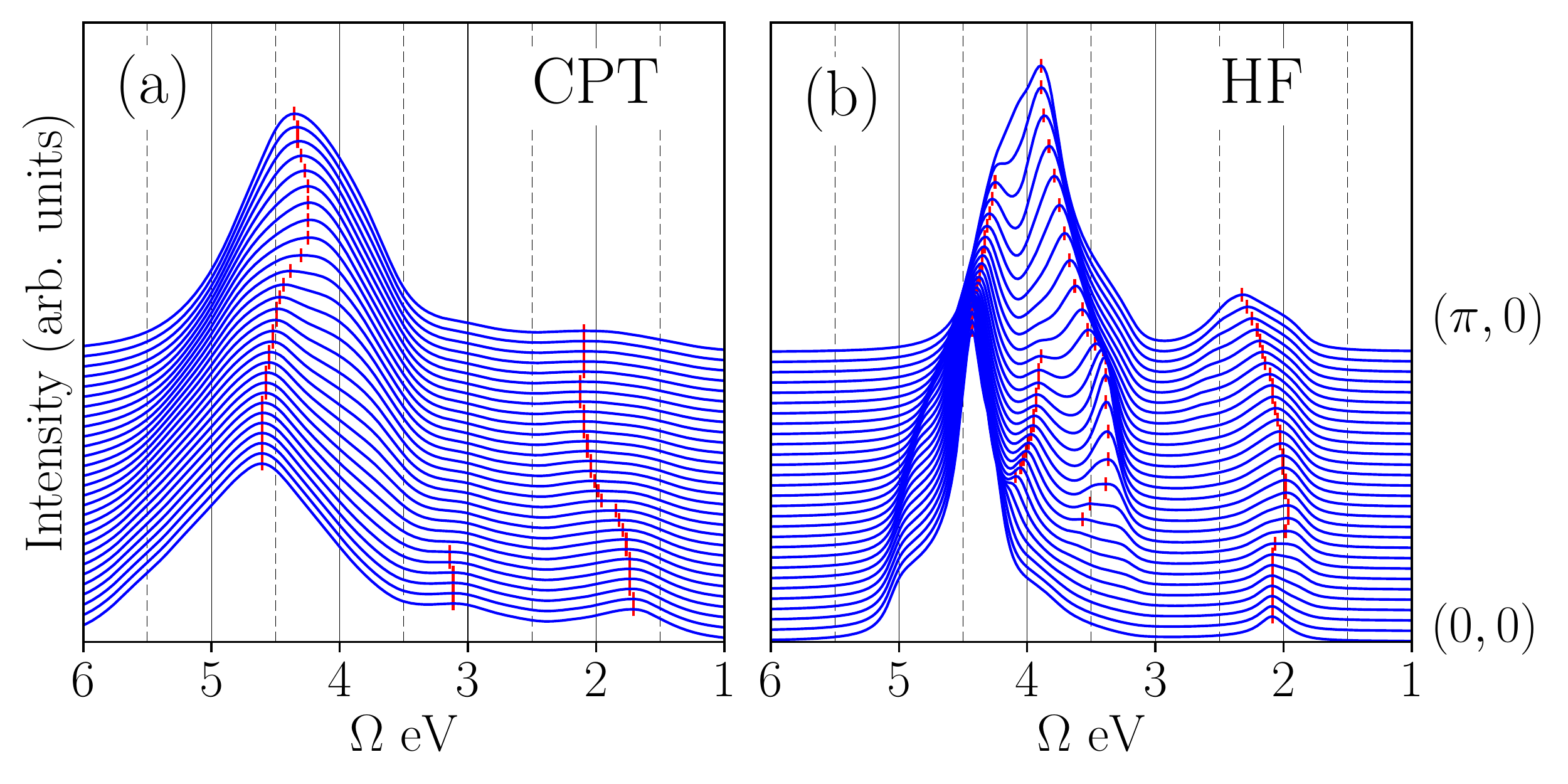}
    \caption{Oxygen $K$-edge RIXS spectra along $\bm{Q}=(0,0)$--$(\pi, 0)$ for (a) CPT and (b) HF.
    Red bars represent positions of peak intensity.
    }
    \label{fig:HFCPT O K-edge RIXS}
\end{figure}
Finally, in order to reveal the effect of the electronic correlation on the RIXS spectra, we examine differences between the two RIXS spectra obtained by CPT and the HFA.
In Fig.~\ref{fig:HFCPT O K-edge RIXS}, we plot the two RIXS spectra along $(0,0)$--$(\pi,0)$, half of which are experimentally accessible momenta.

In Fig.~\ref{fig:HFCPT O K-edge RIXS}(a), the peaks of the ZRS excitation around 2~eV (red bars) shift towards the high energy direction relative to those in Fig.~\ref{fig:HFCPT O K-edge RIXS}(b).
The difference between the momentum dependences obtained by the two calculation schemes is attributed to the difference in between the electronic structures of the O $2p_{x/y}$ orbitals of the ZRS band and the UHB.
This is because the intensity and momentum-dependence of the ZRS excitation is related to the product of the partial occupation number of the O $2p_x$ electrons in the ZRS band and the partial occupation number of the O $2p_x$ holes in the UHB.
The difference can also be attributed to whether the ZRS band is actually a singlet state or not: the ZRS band obtained by CPT is a singlet state described as a many-body state in the range of the reference cluster, whereas the ZRS band obtained by the HFA is an anti-bonding band described as a one-body state.

The difference between the spectra of the CT excitations of the two schemes is in their broadness: the spectra are sharp in the HFA (Fig.~\ref{fig:HFCPT O K-edge RIXS}(b)), while they are broad in CPT (Fig.~\ref{fig:HFCPT O K-edge RIXS}(a)).
One reason for the broad CT excitation spectra is that the self-energy of the Green's functions for the oxygen bands acquires a finite life-time when using CPT.
The other reason is the change in the UHB, which is attributed to the different origins of the UHB.
In the HFA, the charge-transfer gap opens due to folding of the BZ by the AF order.
In CPT, the Coulomb repulsion on the Cu sites opens the energy gap.
We consider that these differences in the oxygen bands and UHB are reflected in the CT excitation. 

%%%%%%%%%%%%%%%%%%%%%%%%%%%%%%%%%%%%%
\section{Summary}
In summary, we calculated the O $K$-edge RIXS spectra for the three-band Hubbard model in the insulating phase by means of three methods.
In particular, we studied elementary charge responses in cuprates includeing the ZRS excitation and CT excitation.

The overall momentum-dependence of the ZRS excitation and CT excitation were revealed by performing a diagrammatic perturbative method in combination with Green's functions obtained by CPT.
The validity of the perturbative method was verified in calculations using the ED method.
Calculations using the ED method indicated that the O $K$-edge RIXS spectra can be interpreted as the dynamical structure factor of O 2$p$ electrons. 
The effect of the electronic correlation on the RIXS spectra was revealed by comparing RIXS spectra obtained by CPT and the HFA.
For example, the peak energies of the ZRS excitation along the $\bm{Q}=(0,0)$--$(\pi, 0)$ obtained by CPT are shifted in the higher energy direction relative to those of the HFA.
Regarding experimental observation of the O $K$-edge RIXS spectra discussed in the present paper, the shift in the momentum-dependence of the ZRS excitation in the high energy direction would be observable in the range of 40$\%$ of the first BZ.

\begin{acknowledgments}
We would like to thank Yusuke Masaki, Hiroaki Matsueda and Atsushi Ono for invaluable discussions.
This work was supported by JST, the establishment of university fellowships towards the creation of science technology innovation, Grant Number JPMJFS2102 and JPSJ KAKENHI Grants No. JP17H02916, No. JP18H05208 and No. JP20H00121.
\end{acknowledgments}

\appendix*
\section{Diagrams in the HFA and CPT}
\label{sec:diagram}

%------------------------------------------------------------------------------
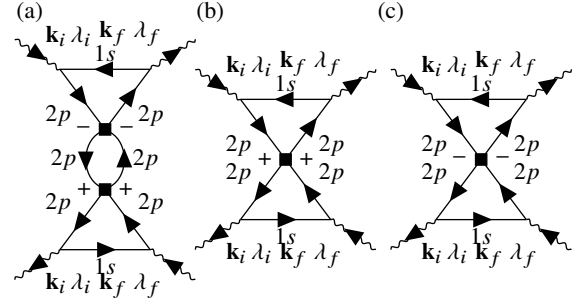
\begin{figure}[t]
\begin{center}
\begin{tikzpicture}[scale=0.4] %tikzpicture環境
\begin{feynhand}    %feynhand環境
    \vertex (a) [label=(a)] at (-2.5, 4) {};
    \vertex (a) [label=(b)] at (-2.5+6, 4) {};
    \vertex (a) [label=(c)] at (-2.5+12, 4) {};
    \vertex [photon] (in)  at (-3,4);
    \vertex [photon] (out) at (3,4);
    \vertex [photon] (rin) at (3,-4);
    \vertex [photon] (rout) at (-3,-4);
    \vertex [particle] (v1) at (-1.5, 3);
    \vertex [particle] (v2) at (1.5, 3);
    \vertex [particle] (v3) at (1.5, -3);
    \vertex [particle] (v4) at (-1.5, -3);
    \vertex[squaredot] (g1) [label=left:$-$, label=right:$-$] at (0,1) {};
    \vertex[squaredot] (g2) [label=left:$+$, label=right:$+$] at (0,-1) {};
    \propag [chabos] (in) to  [edge label=${\bf k}_i \, \lambda_i$] (v1);
    \propag [chabos] (v2) to  [edge label=${\bf k}_f \, \lambda_f$] (out);
    \propag [chabos] (rin) to [edge label=${\bf k}_f \, \lambda_f$] (v3);
    \propag [chabos] (v4) to  [edge label=${\bf k}_i \, \lambda_i$] (rout);
    \propag [anti fermion] (v1) to [edge label=$1s$] (v2);
    \propag [anti fermion] (g1) to [edge label=$2p$] (v1);
    \propag [anti fermion] (v2) to [edge label=$2p$] (g1);
    \propag [anti fermion] (v3) to [edge label=$1s$] (v4);
    \propag [anti fermion] (v4) to [edge label=$2p$] (g2);
    \propag [anti fermion] (g2) to [edge label=$2p$] (v3);
    \propag[fer] (g1) to [out=210, in=150] [edge label'=$2p$] (g2);
    \propag[fer] (g2) to [out=30, in=-30] [edge label'=$2p$] (g1);
    \vertex [photon] (in)  at (-3+6,3);
    \vertex [photon] (out) at (3+6,3);
    \vertex [photon] (rin) at (3+6,-3);
    \vertex [photon] (rout) at (-3+6,-3);
    \vertex [particle] (v1) at (-1.5+6, 2);
    \vertex [particle] (v2) at (1.5+6, 2);
    \vertex [particle] (v3) at (1.5+6, -2);
    \vertex [particle] (v4) at (-1.5+6, -2);
    \vertex[squaredot] (g) [label=left:$+$, label=right:$+$] at (+6,0) {};
    \propag [chabos] (in) to  [edge label=${\bf k}_{i} \, \lambda_i$] (v1);
    \propag [chabos] (v2) to  [edge label=${\bf k}_{f} \, \lambda_f$] (out);
    \propag [chabos] (rin) to [edge label=${\bf k}_{f} \, \lambda_f$] (v3);
    \propag [chabos] (v4) to  [edge label=${\bf k}_{i} \, \lambda_i$] (rout);
    \propag [anti fermion] (v1) to [edge label=$1s$] (v2);
    \propag [anti fermion] (g) to [edge label=$2p$] (v1);
    \propag [anti fermion] (v2) to [edge label=$2p$] (g);
    \propag [anti fermion] (v3) to [edge label=$1s$] (v4);
    \propag [anti fermion] (v4) to [edge label=$2p$] (g);
    \propag [anti fermion] (g) to [edge label=$2p$] (v3);
    \vertex [photon] (in)  at (-3+12.5,3);
    \vertex [photon] (out) at (3+12.5,3);
    \vertex [photon] (rin) at (3+12.5,-3);
    \vertex [photon] (rout) at (-3+12.5,-3);
    \vertex [particle] (v1) at (-1.5+12.5, 2);
    \vertex [particle] (v2) at (1.5+12.5, 2);
    \vertex [particle] (v3) at (1.5+12.5, -2);
    \vertex [particle] (v4) at (-1.5+12.5, -2);
    \vertex[squaredot] (g) [label=left:$-$, label=right:$-$] at (+12.5,0) {};
    \propag [chabos] (in) to  [edge label=${\bf k}_{i} \, \lambda_i$] (v1);
    \propag [chabos] (v2) to  [edge label=${\bf k}_{f} \, \lambda_f$] (out);
    \propag [chabos] (rin) to [edge label=${\bf k}_{f} \, \lambda_f$] (v3);
    \propag [chabos] (v4) to  [edge label=${\bf k}_{i} \, \lambda_i$] (rout);
    \propag [anti fermion] (v1) to [edge label=$1s$] (v2);
    \propag [anti fermion] (g) to [edge label=$2p$] (v1);
    \propag [anti fermion] (v2) to [edge label=$2p$] (g);
    \propag [anti fermion] (v3) to [edge label=$1s$] (v4);
    \propag [anti fermion] (v4) to [edge label=$2p$] (g);
    \propag [anti fermion] (g) to [edge label=$2p$] (v3);
\end{feynhand}
\end{tikzpicture}
\end{center}
\caption{
(a) Diagram for $W_b$ in Eq.~(\ref{eq:wb}). 
Solid lines represent the Green's functions for the $2p$ and $1s$ electrons, and wavy lines represent those for photons. 
Solid squares represent the renormalized vertices. 
(b) and (c) are diagrams for $W_{c}^{\pm}$ in Eq.~(\ref{eq:wc}). 
}
\label{fig:wbc} 
\end{figure}
%------------------------------------------------------------------------------

%------------------------------------------------------------------------------
\begin{figure}[t]
\begin{center}
\begin{tikzpicture}[scale = 0.7] %tikzpicture環境
\begin{feynhand}    %feynhand環境
    \vertex [NEblob] (a) at (0,0) {};
    \vertex [photon] (in1)  at (-3,2.5);
    \vertex [photon] (out1) at (3,2.5);
    \vertex [particle] (v1) [label=above:$t$] at (-1, 2.5);
    \vertex [particle] (v2) [label=above:$0$]at (1, 2.5);
    \vertex [particle] (v3) [label=above:$s$] at (0, 1.5);
    \vertex [photon] (rin1) at (3,-2.5);
    \vertex [photon] (rout1) at (-3,-2.5);
    \vertex [particle] (rv1) [label=below:$u'$] at (1, -2.5);
    \vertex [particle] (rv2) [label=below:$t'$] at (-1, -2.5);
    \vertex [particle] (rv3) [label=below:$s'$] at (0, -1.5);
    \propag [chabos] (in1) to [edge label=$\bm{q}_\tr{i} \, \alpha_{\tr{i}}$] (v1);
    \propag [chabos] (v2) to [edge label=$\bm{q}_\tr{f} \, \alpha_{\tr{f}}$] (out1);
    \propag [fermion] (v1) to [edge label=$2p$] (v2);
    \propag [fermion] (v2) to [edge label=$1s$] (v3);
    \propag [fermion] (v3) to [edge label=$1s$] (v1);
    \propag [chasca] (v3) to [edge label=$V_c$](a);
    \propag [chabos] (rin1) to [edge label=$\bm{q}_\tr{f} \, \alpha_{\tr{f}}$] (rv1);
    \propag [chabos] (rv2) to [edge label=$\bm{q}_\tr{i} \, \alpha_{\tr{i}}$] (rout1);
    \propag [fermion] (rv1) to [edge label=$2p$] (rv2);
    \propag [fermion] (rv2) to [edge label=$1s$] (rv3);
    \propag [fermion] (rv3) to [edge label=$1s$] (rv1);
    \propag [antsca] (rv3) to [edge label=$V_c$](a);
\end{feynhand}
\end{tikzpicture}
\end{center}
\caption{
Diagram for $W_{\mathrm{indirect}}$ in Eq.~(\ref{eq:windirect}). 
Solid lines represent the Green's functions for the $2p$ and $1s$ electrons, and wavy lines represent those for photons, and dotted lines represent the Coulomb interaction. 
The shaded circle represents the density–density correlation function of the Keldysh-type.
}
\label{fig:windirect} 
\end{figure}
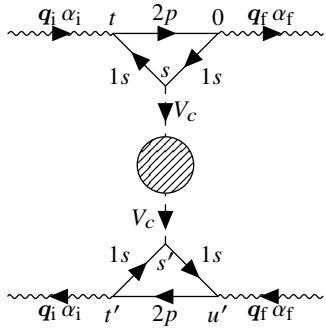
%------------------------------------------------------------------------------

In this Appendix, we show the explicit forms of the RIXS diagrams. 
In addition to the $W_a$ diagram shown in Fig.~\ref{fig:wa}, we consider  
three other diagrams: $W_b$, $W_c$ and $W_{\mathrm{indirect}}$. 
First, we give the explicit form of the diagram $W_b$ shown in Fig.~\ref{fig:wbc}(a): 
\begin{align}
W_b&=
|w_{\bm{k}_f \lambda_f} w_{\bm{k}_i \lambda_i }|^2
\frac{1}{N} \sum_{\sigma \sigma'} \sum_{\alpha \alpha'}  e_{\lambda_i}^{\alpha}  e_{\lambda_f}^{\alpha}  e_{\lambda_i}^{\alpha'} e_{\lambda_f}^{\alpha'} 
\nonumber \\
&\times
\left[
L(\omega_i, \omega_f, \bm{Q})^\dagger\Gamma_0\Gamma(\bm{Q})^\dagger\Pi_{0}^{+-}(\bm{Q}) 
\right. 
\nonumber \\
& \times \left.
\Gamma(Q)\Gamma_0 L(\omega_i, \omega_f, \bm{Q})
\right]_{\alpha'\sigma'; \alpha\sigma} . 
\label{eq:wb}
\end{align}
Here, $\Pi_{\alpha'\sigma';\alpha\sigma}^{(0)+-}(\bm{Q})$ is the bubble part of the diagram given by 
\begin{align}
& \Pi_{\alpha'\sigma';\alpha\sigma}^{(0)+-}(\bm{Q})  =-i\dfrac{1}{N}\sum_{\bm{k}}
 \int_{-\infty}^{\infty}\dfrac{d\omega}{2\pi}
       \nonumber \\
   &  \ \ \ \times G^{2p+-}_{\alpha'\sigma'; \alpha\sigma}(\bm{k}+\bm{Q}, \omega+\Omega)
    G^{2p-+}_{\alpha\sigma; \alpha'\sigma'}(\bm{k}, \omega) , 
\end{align}
and $L(\omega_i, \omega_f, \bm{Q})$ is the triangle part given by 
\begin{equation}
\begin{split}
&
    \left[L(\omega_i, \omega_f, \bm{Q})\right]_{\alpha'\sigma' ;\alpha\sigma}\\
    &=
    \dfrac{1}{N}\sum_{\bm{k}}
    \int\dfrac{d\omega_1 d\omega_2}{(2\pi)^2}
    \left[
    \dfrac{R(\omega_i, \omega_2)}{\Omega + \omega_1 - \omega_2 + i\eta}
    G^{2p-+}_{\alpha'\sigma';\alpha\sigma}(\bm{k} + \bm{Q}, \omega_1)
    \right. \\& \times \left.
    G^{2p+-}_{\alpha\sigma;\alpha'\sigma'}(\bm{k}, \omega_2)
    +
    \dfrac{R(\omega_f, \omega_1)}{\Omega + \omega_1 - \omega_2 - i\eta}
    G^{2p+-}_{\alpha' \sigma';\alpha \sigma}(\bm{k} + \bm{Q}, \omega_1)
    \right. \\& \times \left.
    G^{2p-+}_{\alpha \sigma;\alpha '\sigma'}(\bm{k}, \omega_2)
    -
    R(\omega_i, \omega_2)
    R(\omega_f, \omega_1)
    \right.\\& \times \left.
    G^{2p+-}_{\alpha'\sigma';\alpha \sigma}(\bm{k} + \bm{Q}, \omega_1)
    G^{2p+-}_{\alpha\sigma;\alpha' \sigma'}(\bm{k}, \omega_2)
    \right]
\end{split} , 
\end{equation}
where $R(\omega_1, \omega_2)$ is introduced in Eq.~(\ref{eq:Rxy}).
The bare vertex for the Coulomb interaction is defined as 
$[\Gamma_0]_{\alpha, \sigma, \alpha', \sigma'}=U^{p}\delta_{\alpha, \alpha'}(1-\delta_{\sigma, \sigma'})$ 
and the renormalized vertex 
$\Gamma(Q)=[I-\Gamma^0 \Pi^{(0)}(Q)]^{-1}$. 
Next, $W_c$ is divided into two parts: 
\begin{align}
&
W_c = W_c^{-}+W_c^{+}\\
&=
|w_{\bm{k}_f \lambda_f} w_{\bm{k}_i \lambda_i }|^2
\sum_{\sigma \sigma'} \sum_{\alpha \alpha'}  e_{\lambda_i}^{\alpha}  e_{\lambda_f}^{\alpha}  e_{\lambda_i}^{\alpha'} e_{\lambda_f}^{\alpha'} 
\nonumber \\
&\times
\left[
N(\omega_i, Q)^\dagger \Gamma(Q)^\dagger \Gamma_0 L(\omega_i, \omega_f, Q)
\right. \nonumber \\ 
&+\left.
L(\omega_i, \omega_f, Q)^\dagger \Gamma_0 \Gamma(Q)^\dagger N(\omega_i, Q)
\right]_{\alpha'\sigma';\alpha\sigma} , 
\label{eq:wc}
\end{align}
with the triangle part defined by 
\begin{align}
&
\left[N(\omega_i, \omega_f, Q)\right]_{\alpha'\sigma' ;\alpha\sigma}
\nonumber \\
&=
\frac{1}{N} \sum_{\bm{k}}
\int \dfrac{d\omega}{2\pi}
R(\omega_i, \omega + \Omega)
\nonumber \\
&
G^{2p+-}_{\alpha'\sigma' ; \alpha\sigma}(\bm{k}+\bm{Q}, \omega + \Omega)
G^{2p-+}_{\alpha\sigma ; \alpha'\sigma'}(\bm{k},\omega) .
\end{align}
Finally, $W_{\mathrm{indirect}}$ in Fig.~\ref{fig:windirect} is given by 
\begin{align}
W_{\mathrm{indirect}}&=
|w_{\bm{k}_f \lambda_f} w_{\bm{k}_i \lambda_i }|^2
\frac{1}{N} \sum_{\bm{k} \sigma \sigma'} \sum_{\alpha \alpha'}  e_{\lambda_i}^{\alpha}  e_{\lambda_f}^{\alpha}  e_{\lambda_i}^{\alpha'}  e_{\lambda_f}^{\alpha'}
\nonumber \\
&
\times
\left[\Gamma^\dagger(\bm{Q})\Pi^{+-}_0(\bm{Q})\Gamma(\bm{Q})\right]_{\alpha'\sigma';\alpha\sigma}
\left|M(\omega_i, \Omega)\right|^2_{\alpha'\sigma';\alpha\sigma}
\label{eq:windirect}
\end{align}
with
\begin{align}
&
    \left[M(\omega_{i}, \Omega)\right]_{\alpha'\sigma' ;\alpha\sigma}
\nonumber \\
    &=
    \dfrac{V_c}{N}
    \sum_{\bm{k}}
    \int\dfrac{d\omega}{2\pi}
    R(\omega_i, \omega)R(\omega_i, \omega + \Omega)
    G^{2p+-}_{\alpha'\sigma'; \alpha\sigma}(\bm{k}, \omega).
\end{align}

%------------------------------------------------------------------------------
%\begin{figure}[t]
%\begin{center}
%\includegraphics[width=1.0\columnwidth,clip]{Figures/bubble}
%\end{center}
%\caption{
%A diagram for $W_{indirect}$ given in Eq.~(\ref{eq:bubble}). 
%Solid lines, wavy dotted lines represent the Green's functions for the $2p$ and $1s$ electrons, those for photon, 
%and the Coulomb interaction, respectively. 
%A shaded circles is the 
%}
%\label{fig:bubble} 
%\end{figure}
%------------------------------------------------------------------------------
$W_{\mathrm{indirect}}$ corresponds to the lowest order contribution expanded for the Coulomb attraction between the valence electron and core hole, the details of which are given in Ref.~\cite{Igarashi_2006_K-edge2}.

\bibliographystyle{apsrev4-1}
\bibliography{bibl}

\begin{thebibliography}{45}%
\makeatletter
\providecommand \@ifxundefined [1]{%
 \@ifx{#1\undefined}
}%
\providecommand \@ifnum [1]{%
 \ifnum #1\expandafter \@firstoftwo
 \else \expandafter \@secondoftwo
 \fi
}%
\providecommand \@ifx [1]{%
 \ifx #1\expandafter \@firstoftwo
 \else \expandafter \@secondoftwo
 \fi
}%
\providecommand \natexlab [1]{#1}%
\providecommand \enquote  [1]{``#1''}%
\providecommand \bibnamefont  [1]{#1}%
\providecommand \bibfnamefont [1]{#1}%
\providecommand \citenamefont [1]{#1}%
\providecommand \href@noop [0]{\@secondoftwo}%
\providecommand \href [0]{\begingroup \@sanitize@url \@href}%
\providecommand \@href[1]{\@@startlink{#1}\@@href}%
\providecommand \@@href[1]{\endgroup#1\@@endlink}%
\providecommand \@sanitize@url [0]{\catcode `\\12\catcode `\$12\catcode
  `\&12\catcode `\#12\catcode `\^12\catcode `\_12\catcode `\%12\relax}%
\providecommand \@@startlink[1]{}%
\providecommand \@@endlink[0]{}%
\providecommand \url  [0]{\begingroup\@sanitize@url \@url }%
\providecommand \@url [1]{\endgroup\@href {#1}{\urlprefix }}%
\providecommand \urlprefix  [0]{URL }%
\providecommand \Eprint [0]{\href }%
\providecommand \doibase [0]{http://dx.doi.org/}%
\providecommand \selectlanguage [0]{\@gobble}%
\providecommand \bibinfo  [0]{\@secondoftwo}%
\providecommand \bibfield  [0]{\@secondoftwo}%
\providecommand \translation [1]{[#1]}%
\providecommand \BibitemOpen [0]{}%
\providecommand \bibitemStop [0]{}%
\providecommand \bibitemNoStop [0]{.\EOS\space}%
\providecommand \EOS [0]{\spacefactor3000\relax}%
\providecommand \BibitemShut  [1]{\csname bibitem#1\endcsname}%
\let\auto@bib@innerbib\@empty
%</preamble>
\bibitem [{\citenamefont {Kotani}\ and\ \citenamefont
  {Shin}(2001)}]{Kotani_Shin_2001_review}%
  \BibitemOpen
  \bibfield  {author} {\bibinfo {author} {\bibfnamefont {A.}~\bibnamefont
  {Kotani}}\ and\ \bibinfo {author} {\bibfnamefont {S.}~\bibnamefont {Shin}},\
  }\href {\doibase 10.1103/RevModPhys.73.203} {\bibfield  {journal} {\bibinfo
  {journal} {Rev. Mod. Phys.}\ }\textbf {\bibinfo {volume} {73}},\ \bibinfo
  {pages} {203} (\bibinfo {year} {2001})}\BibitemShut {NoStop}%
\bibitem [{\citenamefont {Ament}\ \emph {et~al.}(2011)\citenamefont {Ament},
  \citenamefont {van Veenendaal}, \citenamefont {Devereaux}, \citenamefont
  {Hill},\ and\ \citenamefont {van~den Brink}}]{Ament_2011_review}%
  \BibitemOpen
  \bibfield  {author} {\bibinfo {author} {\bibfnamefont {L.~J.~P.}\
  \bibnamefont {Ament}}, \bibinfo {author} {\bibfnamefont {M.}~\bibnamefont
  {van Veenendaal}}, \bibinfo {author} {\bibfnamefont {T.~P.}\ \bibnamefont
  {Devereaux}}, \bibinfo {author} {\bibfnamefont {J.~P.}\ \bibnamefont {Hill}},
  \ and\ \bibinfo {author} {\bibfnamefont {J.}~\bibnamefont {van~den Brink}},\
  }\href {\doibase 10.1103/RevModPhys.83.705} {\bibfield  {journal} {\bibinfo
  {journal} {Rev. Mod. Phys.}\ }\textbf {\bibinfo {volume} {83}},\ \bibinfo
  {pages} {705} (\bibinfo {year} {2011})}\BibitemShut {NoStop}%
\bibitem [{\citenamefont {Ishii}\ \emph {et~al.}(2013)\citenamefont {Ishii},
  \citenamefont {Tohyama},\ and\ \citenamefont
  {Mizuki}}]{IshiiTohyama_2013_review}%
  \BibitemOpen
  \bibfield  {author} {\bibinfo {author} {\bibfnamefont {K.}~\bibnamefont
  {Ishii}}, \bibinfo {author} {\bibfnamefont {T.}~\bibnamefont {Tohyama}}, \
  and\ \bibinfo {author} {\bibfnamefont {J.}~\bibnamefont {Mizuki}},\ }\href
  {\doibase 10.7566/JPSJ.82.021015} {\bibfield  {journal} {\bibinfo  {journal}
  {J. Phys. Soc. Jpn.}\ }\textbf {\bibinfo {volume} {82}},\ \bibinfo {pages}
  {021015} (\bibinfo {year} {2013})}\BibitemShut {NoStop}%
\bibitem [{\citenamefont {Wang}\ \emph {et~al.}(2018)\citenamefont {Wang},
  \citenamefont {Claassen}, \citenamefont {Pemmaraju}, \citenamefont {Jia},
  \citenamefont {Moritz},\ and\ \citenamefont
  {Devereaux}}]{Yao_2018_RIXSReviewNature}%
  \BibitemOpen
  \bibfield  {author} {\bibinfo {author} {\bibfnamefont {Y.}~\bibnamefont
  {Wang}}, \bibinfo {author} {\bibfnamefont {M.}~\bibnamefont {Claassen}},
  \bibinfo {author} {\bibfnamefont {C.~D.}\ \bibnamefont {Pemmaraju}}, \bibinfo
  {author} {\bibfnamefont {C.}~\bibnamefont {Jia}}, \bibinfo {author}
  {\bibfnamefont {B.}~\bibnamefont {Moritz}}, \ and\ \bibinfo {author}
  {\bibfnamefont {T.~P.}\ \bibnamefont {Devereaux}},\ }\href {\doibase
  10.1038/s41578-018-0046-3} {\bibfield  {journal} {\bibinfo  {journal} {Nat.
  Rev. Mater.}\ }\textbf {\bibinfo {volume} {3}},\ \bibinfo {pages} {312}
  (\bibinfo {year} {2018})}\BibitemShut {NoStop}%
\bibitem [{\citenamefont {Suga}\ \emph {et~al.}(2005)\citenamefont {Suga},
  \citenamefont {Imada}, \citenamefont {Higashiya}, \citenamefont {Shigemoto},
  \citenamefont {Kasai}, \citenamefont {Sing}, \citenamefont {Fujiwara},
  \citenamefont {Sekiyama}, \citenamefont {Yamasaki}, \citenamefont {Kim},
  \citenamefont {Nomura}, \citenamefont {Igarashi}, \citenamefont {Yabashi},\
  and\ \citenamefont {Ishikawa}}]{Suga2005}%
  \BibitemOpen
  \bibfield  {author} {\bibinfo {author} {\bibfnamefont {S.}~\bibnamefont
  {Suga}}, \bibinfo {author} {\bibfnamefont {S.}~\bibnamefont {Imada}},
  \bibinfo {author} {\bibfnamefont {A.}~\bibnamefont {Higashiya}}, \bibinfo
  {author} {\bibfnamefont {A.}~\bibnamefont {Shigemoto}}, \bibinfo {author}
  {\bibfnamefont {S.}~\bibnamefont {Kasai}}, \bibinfo {author} {\bibfnamefont
  {M.}~\bibnamefont {Sing}}, \bibinfo {author} {\bibfnamefont {H.}~\bibnamefont
  {Fujiwara}}, \bibinfo {author} {\bibfnamefont {A.}~\bibnamefont {Sekiyama}},
  \bibinfo {author} {\bibfnamefont {A.}~\bibnamefont {Yamasaki}}, \bibinfo
  {author} {\bibfnamefont {C.}~\bibnamefont {Kim}}, \bibinfo {author}
  {\bibfnamefont {T.}~\bibnamefont {Nomura}}, \bibinfo {author} {\bibfnamefont
  {J.}~\bibnamefont {Igarashi}}, \bibinfo {author} {\bibfnamefont
  {M.}~\bibnamefont {Yabashi}}, \ and\ \bibinfo {author} {\bibfnamefont
  {T.}~\bibnamefont {Ishikawa}},\ }\href {\doibase 10.1103/PhysRevB.72.081101}
  {\bibfield  {journal} {\bibinfo  {journal} {Phys. Rev. B}\ }\textbf {\bibinfo
  {volume} {72}},\ \bibinfo {pages} {081101} (\bibinfo {year}
  {2005})}\BibitemShut {NoStop}%
\bibitem [{\citenamefont {Marra}\ \emph {et~al.}(2013)\citenamefont {Marra},
  \citenamefont {Sykora}, \citenamefont {Wohlfeld},\ and\ \citenamefont {{Van
  Den Brink}}}]{Marra2013}%
  \BibitemOpen
  \bibfield  {author} {\bibinfo {author} {\bibfnamefont {P.}~\bibnamefont
  {Marra}}, \bibinfo {author} {\bibfnamefont {S.}~\bibnamefont {Sykora}},
  \bibinfo {author} {\bibfnamefont {K.}~\bibnamefont {Wohlfeld}}, \ and\
  \bibinfo {author} {\bibfnamefont {J.}~\bibnamefont {{Van Den Brink}}},\
  }\href {\doibase 10.1103/PhysRevLett.110.117005} {\bibfield  {journal}
  {\bibinfo  {journal} {Phys. Rev. Lett.}\ }\textbf {\bibinfo {volume} {110}},\
  \bibinfo {pages} {1} (\bibinfo {year} {2013})}\BibitemShut {NoStop}%
\bibitem [{\citenamefont {Marra}\ \emph {et~al.}(2016)\citenamefont {Marra},
  \citenamefont {{Van Den Brink}},\ and\ \citenamefont {Sykora}}]{Marra2016}%
  \BibitemOpen
  \bibfield  {author} {\bibinfo {author} {\bibfnamefont {P.}~\bibnamefont
  {Marra}}, \bibinfo {author} {\bibfnamefont {J.}~\bibnamefont {{Van Den
  Brink}}}, \ and\ \bibinfo {author} {\bibfnamefont {S.}~\bibnamefont
  {Sykora}},\ }\href {\doibase 10.1038/srep25386} {\bibfield  {journal}
  {\bibinfo  {journal} {Sci. Rep.}\ }\textbf {\bibinfo {volume} {6}},\ \bibinfo
  {pages} {1} (\bibinfo {year} {2016})},\ \Eprint
  {http://arxiv.org/abs/1405.5556} {1405.5556} \BibitemShut {NoStop}%
\bibitem [{\citenamefont {Ament}\ \emph {et~al.}(2007)\citenamefont {Ament},
  \citenamefont {Forte},\ and\ \citenamefont {{Van Den Brink}}}]{Ament2007}%
  \BibitemOpen
  \bibfield  {author} {\bibinfo {author} {\bibfnamefont {L.~J.}\ \bibnamefont
  {Ament}}, \bibinfo {author} {\bibfnamefont {F.}~\bibnamefont {Forte}}, \ and\
  \bibinfo {author} {\bibfnamefont {J.}~\bibnamefont {{Van Den Brink}}},\
  }\href@noop {} {\bibfield  {journal} {\bibinfo  {journal} {Phys. Rev. B}\
  }\textbf {\bibinfo {volume} {75}} (\bibinfo {year} {2007})}\BibitemShut
  {NoStop}%
\bibitem [{\citenamefont {Tsutsui}\ and\ \citenamefont
  {Tohyama}(2016)}]{Tsutsui2016}%
  \BibitemOpen
  \bibfield  {author} {\bibinfo {author} {\bibfnamefont {K.}~\bibnamefont
  {Tsutsui}}\ and\ \bibinfo {author} {\bibfnamefont {T.}~\bibnamefont
  {Tohyama}},\ }\href {\doibase 10.1103/PhysRevB.94.085144} {\bibfield
  {journal} {\bibinfo  {journal} {Phys. Rev. B}\ }\textbf {\bibinfo {volume}
  {94}},\ \bibinfo {pages} {085144} (\bibinfo {year} {2016})}\BibitemShut
  {NoStop}%
\bibitem [{\citenamefont {Jia}\ \emph {et~al.}(2016)\citenamefont {Jia},
  \citenamefont {Wohlfeld}, \citenamefont {Wang}, \citenamefont {Moritz},\ and\
  \citenamefont {Devereaux}}]{Jia2016}%
  \BibitemOpen
  \bibfield  {author} {\bibinfo {author} {\bibfnamefont {C.}~\bibnamefont
  {Jia}}, \bibinfo {author} {\bibfnamefont {K.}~\bibnamefont {Wohlfeld}},
  \bibinfo {author} {\bibfnamefont {Y.}~\bibnamefont {Wang}}, \bibinfo {author}
  {\bibfnamefont {B.}~\bibnamefont {Moritz}}, \ and\ \bibinfo {author}
  {\bibfnamefont {T.~P.}\ \bibnamefont {Devereaux}},\ }\href {\doibase
  10.1103/PhysRevX.6.021020} {\bibfield  {journal} {\bibinfo  {journal} {Phys.
  Rev. X}\ }\textbf {\bibinfo {volume} {6}},\ \bibinfo {pages} {021020}
  (\bibinfo {year} {2016})}\BibitemShut {NoStop}%
\bibitem [{\citenamefont {Nocera}\ \emph {et~al.}(2018)\citenamefont {Nocera},
  \citenamefont {Kumar}, \citenamefont {Kaushal}, \citenamefont {Alvarez},
  \citenamefont {Dagotto},\ and\ \citenamefont {Johnston}}]{Nocera2018}%
  \BibitemOpen
  \bibfield  {author} {\bibinfo {author} {\bibfnamefont {A.}~\bibnamefont
  {Nocera}}, \bibinfo {author} {\bibfnamefont {U.}~\bibnamefont {Kumar}},
  \bibinfo {author} {\bibfnamefont {N.}~\bibnamefont {Kaushal}}, \bibinfo
  {author} {\bibfnamefont {G.}~\bibnamefont {Alvarez}}, \bibinfo {author}
  {\bibfnamefont {E.}~\bibnamefont {Dagotto}}, \ and\ \bibinfo {author}
  {\bibfnamefont {S.}~\bibnamefont {Johnston}},\ }\href {\doibase
  10.1038/s41598-018-29218-8} {\bibfield  {journal} {\bibinfo  {journal} {Sci.
  Rep.}\ }\textbf {\bibinfo {volume} {8}},\ \bibinfo {pages} {1} (\bibinfo
  {year} {2018})}\BibitemShut {NoStop}%
\bibitem [{\citenamefont {Zhang}\ and\ \citenamefont
  {Rice}(1988)}]{Zhang_Rice_1988}%
  \BibitemOpen
  \bibfield  {author} {\bibinfo {author} {\bibfnamefont {F.~C.}\ \bibnamefont
  {Zhang}}\ and\ \bibinfo {author} {\bibfnamefont {T.~M.}\ \bibnamefont
  {Rice}},\ }\href {\doibase 10.1103/PhysRevB.37.3759} {\bibfield  {journal}
  {\bibinfo  {journal} {Phys. Rev. B}\ }\textbf {\bibinfo {volume} {37}},\
  \bibinfo {pages} {3759} (\bibinfo {year} {1988})}\BibitemShut {NoStop}%
\bibitem [{\citenamefont {Chen}\ \emph
  {et~al.}(2013{\natexlab{a}})\citenamefont {Chen}, \citenamefont {Jiang},
  \citenamefont {Luo}, \citenamefont {Lin}, \citenamefont {Wu}, \citenamefont
  {Lee}, \citenamefont {Chen}, \citenamefont {Kuo}, \citenamefont {Juang},\
  and\ \citenamefont {Mou}}]{Chen_2013_ZRS_XAS}%
  \BibitemOpen
  \bibfield  {author} {\bibinfo {author} {\bibfnamefont {Y.-J.}\ \bibnamefont
  {Chen}}, \bibinfo {author} {\bibfnamefont {M.~G.}\ \bibnamefont {Jiang}},
  \bibinfo {author} {\bibfnamefont {C.~W.}\ \bibnamefont {Luo}}, \bibinfo
  {author} {\bibfnamefont {J.-Y.}\ \bibnamefont {Lin}}, \bibinfo {author}
  {\bibfnamefont {K.~H.}\ \bibnamefont {Wu}}, \bibinfo {author} {\bibfnamefont
  {J.~M.}\ \bibnamefont {Lee}}, \bibinfo {author} {\bibfnamefont {J.~M.}\
  \bibnamefont {Chen}}, \bibinfo {author} {\bibfnamefont {Y.~K.}\ \bibnamefont
  {Kuo}}, \bibinfo {author} {\bibfnamefont {J.~Y.}\ \bibnamefont {Juang}}, \
  and\ \bibinfo {author} {\bibfnamefont {C.-Y.}\ \bibnamefont {Mou}},\ }\href
  {\doibase 10.1103/PhysRevB.88.134525} {\bibfield  {journal} {\bibinfo
  {journal} {Phys. Rev. B}\ }\textbf {\bibinfo {volume} {88}},\ \bibinfo
  {pages} {134525} (\bibinfo {year} {2013}{\natexlab{a}})}\BibitemShut
  {NoStop}%
\bibitem [{\citenamefont {Chen}\ \emph
  {et~al.}(2013{\natexlab{b}})\citenamefont {Chen}, \citenamefont {Sentef},
  \citenamefont {Kung}, \citenamefont {Jia}, \citenamefont {Thomale},
  \citenamefont {Moritz}, \citenamefont {Kampf},\ and\ \citenamefont
  {Devereaux}}]{Chen_2013_ZRS_OXAS}%
  \BibitemOpen
  \bibfield  {author} {\bibinfo {author} {\bibfnamefont {C.-C.}\ \bibnamefont
  {Chen}}, \bibinfo {author} {\bibfnamefont {M.}~\bibnamefont {Sentef}},
  \bibinfo {author} {\bibfnamefont {Y.~F.}\ \bibnamefont {Kung}}, \bibinfo
  {author} {\bibfnamefont {C.~J.}\ \bibnamefont {Jia}}, \bibinfo {author}
  {\bibfnamefont {R.}~\bibnamefont {Thomale}}, \bibinfo {author} {\bibfnamefont
  {B.}~\bibnamefont {Moritz}}, \bibinfo {author} {\bibfnamefont {A.~P.}\
  \bibnamefont {Kampf}}, \ and\ \bibinfo {author} {\bibfnamefont {T.~P.}\
  \bibnamefont {Devereaux}},\ }\href {\doibase 10.1103/PhysRevB.87.165144}
  {\bibfield  {journal} {\bibinfo  {journal} {Phys. Rev. B}\ }\textbf {\bibinfo
  {volume} {87}},\ \bibinfo {pages} {165144} (\bibinfo {year}
  {2013}{\natexlab{b}})}\BibitemShut {NoStop}%
\bibitem [{\citenamefont {Kung}\ \emph {et~al.}(2016)\citenamefont {Kung},
  \citenamefont {Chen}, \citenamefont {Wang}, \citenamefont {Huang},
  \citenamefont {Nowadnick}, \citenamefont {Moritz}, \citenamefont {Scalettar},
  \citenamefont {Johnston},\ and\ \citenamefont {Devereaux}}]{Kung_2016_TBHBM}%
  \BibitemOpen
  \bibfield  {author} {\bibinfo {author} {\bibfnamefont {Y.~F.}\ \bibnamefont
  {Kung}}, \bibinfo {author} {\bibfnamefont {C.-C.}\ \bibnamefont {Chen}},
  \bibinfo {author} {\bibfnamefont {Y.}~\bibnamefont {Wang}}, \bibinfo {author}
  {\bibfnamefont {E.~W.}\ \bibnamefont {Huang}}, \bibinfo {author}
  {\bibfnamefont {E.~A.}\ \bibnamefont {Nowadnick}}, \bibinfo {author}
  {\bibfnamefont {B.}~\bibnamefont {Moritz}}, \bibinfo {author} {\bibfnamefont
  {R.~T.}\ \bibnamefont {Scalettar}}, \bibinfo {author} {\bibfnamefont
  {S.}~\bibnamefont {Johnston}}, \ and\ \bibinfo {author} {\bibfnamefont
  {T.~P.}\ \bibnamefont {Devereaux}},\ }\href {\doibase
  10.1103/PhysRevB.93.155166} {\bibfield  {journal} {\bibinfo  {journal} {Phys.
  Rev. B}\ }\textbf {\bibinfo {volume} {93}},\ \bibinfo {pages} {155166}
  (\bibinfo {year} {2016})}\BibitemShut {NoStop}%
\bibitem [{\citenamefont {Monney}\ \emph {et~al.}(2016)\citenamefont {Monney},
  \citenamefont {Bisogni}, \citenamefont {Zhou}, \citenamefont {Kraus},
  \citenamefont {Strocov}, \citenamefont {Behr}, \citenamefont {Drechsler},
  \citenamefont {Rosner}, \citenamefont {Johnston}, \citenamefont {Geck},\ and\
  \citenamefont {Schmitt}}]{Monney_2016_ZRS}%
  \BibitemOpen
  \bibfield  {author} {\bibinfo {author} {\bibfnamefont {C.}~\bibnamefont
  {Monney}}, \bibinfo {author} {\bibfnamefont {V.}~\bibnamefont {Bisogni}},
  \bibinfo {author} {\bibfnamefont {K.-J.}\ \bibnamefont {Zhou}}, \bibinfo
  {author} {\bibfnamefont {R.}~\bibnamefont {Kraus}}, \bibinfo {author}
  {\bibfnamefont {V.~N.}\ \bibnamefont {Strocov}}, \bibinfo {author}
  {\bibfnamefont {G.}~\bibnamefont {Behr}}, \bibinfo {author} {\bibfnamefont
  {S.-L.}\ \bibnamefont {Drechsler}}, \bibinfo {author} {\bibfnamefont
  {H.}~\bibnamefont {Rosner}}, \bibinfo {author} {\bibfnamefont
  {S.}~\bibnamefont {Johnston}}, \bibinfo {author} {\bibfnamefont
  {J.}~\bibnamefont {Geck}}, \ and\ \bibinfo {author} {\bibfnamefont
  {T.}~\bibnamefont {Schmitt}},\ }\href {\doibase 10.1103/PhysRevB.94.165118}
  {\bibfield  {journal} {\bibinfo  {journal} {Phys. Rev. B}\ }\textbf {\bibinfo
  {volume} {94}},\ \bibinfo {pages} {165118} (\bibinfo {year}
  {2016})}\BibitemShut {NoStop}%
\bibitem [{\citenamefont {Kim}\ \emph {et~al.}(2002)\citenamefont {Kim},
  \citenamefont {Hill}, \citenamefont {Burns}, \citenamefont {Wakimoto},
  \citenamefont {Birgeneau}, \citenamefont {Casa}, \citenamefont {Gog},\ and\
  \citenamefont {Venkataraman}}]{Kim_2002_CuKexp}%
  \BibitemOpen
  \bibfield  {author} {\bibinfo {author} {\bibfnamefont {Y.~J.}\ \bibnamefont
  {Kim}}, \bibinfo {author} {\bibfnamefont {J.~P.}\ \bibnamefont {Hill}},
  \bibinfo {author} {\bibfnamefont {C.~A.}\ \bibnamefont {Burns}}, \bibinfo
  {author} {\bibfnamefont {S.}~\bibnamefont {Wakimoto}}, \bibinfo {author}
  {\bibfnamefont {R.~J.}\ \bibnamefont {Birgeneau}}, \bibinfo {author}
  {\bibfnamefont {D.}~\bibnamefont {Casa}}, \bibinfo {author} {\bibfnamefont
  {T.}~\bibnamefont {Gog}}, \ and\ \bibinfo {author} {\bibfnamefont {C.~T.}\
  \bibnamefont {Venkataraman}},\ }\href {\doibase
  10.1103/PhysRevLett.89.177003} {\bibfield  {journal} {\bibinfo  {journal}
  {Phys. Rev. Lett.}\ }\textbf {\bibinfo {volume} {89}},\ \bibinfo {pages}
  {177003} (\bibinfo {year} {2002})}\BibitemShut {NoStop}%
\bibitem [{\citenamefont {Ellis}\ \emph {et~al.}(2008)\citenamefont {Ellis},
  \citenamefont {Hill}, \citenamefont {Wakimoto}, \citenamefont {Birgeneau},
  \citenamefont {Casa}, \citenamefont {Gog},\ and\ \citenamefont
  {Kim}}]{Ellis_2008_LCOexp}%
  \BibitemOpen
  \bibfield  {author} {\bibinfo {author} {\bibfnamefont {D.~S.}\ \bibnamefont
  {Ellis}}, \bibinfo {author} {\bibfnamefont {J.~P.}\ \bibnamefont {Hill}},
  \bibinfo {author} {\bibfnamefont {S.}~\bibnamefont {Wakimoto}}, \bibinfo
  {author} {\bibfnamefont {R.~J.}\ \bibnamefont {Birgeneau}}, \bibinfo {author}
  {\bibfnamefont {D.}~\bibnamefont {Casa}}, \bibinfo {author} {\bibfnamefont
  {T.}~\bibnamefont {Gog}}, \ and\ \bibinfo {author} {\bibfnamefont {Y.-J.}\
  \bibnamefont {Kim}},\ }\href {\doibase 10.1103/PhysRevB.77.060501} {\bibfield
   {journal} {\bibinfo  {journal} {Phys. Rev. B}\ }\textbf {\bibinfo {volume}
  {77}},\ \bibinfo {pages} {060501} (\bibinfo {year} {2008})}\BibitemShut
  {NoStop}%
\bibitem [{\citenamefont {Ellis}\ \emph {et~al.}(2011)\citenamefont {Ellis},
  \citenamefont {Kim}, \citenamefont {Zhang}, \citenamefont {Hill},
  \citenamefont {Gu}, \citenamefont {Komiya}, \citenamefont {Ando},
  \citenamefont {Casa}, \citenamefont {Gog},\ and\ \citenamefont
  {Kim}}]{Ellis_2011_CuK}%
  \BibitemOpen
  \bibfield  {author} {\bibinfo {author} {\bibfnamefont {D.~S.}\ \bibnamefont
  {Ellis}}, \bibinfo {author} {\bibfnamefont {J.}~\bibnamefont {Kim}}, \bibinfo
  {author} {\bibfnamefont {H.}~\bibnamefont {Zhang}}, \bibinfo {author}
  {\bibfnamefont {J.~P.}\ \bibnamefont {Hill}}, \bibinfo {author}
  {\bibfnamefont {G.}~\bibnamefont {Gu}}, \bibinfo {author} {\bibfnamefont
  {S.}~\bibnamefont {Komiya}}, \bibinfo {author} {\bibfnamefont
  {Y.}~\bibnamefont {Ando}}, \bibinfo {author} {\bibfnamefont {D.}~\bibnamefont
  {Casa}}, \bibinfo {author} {\bibfnamefont {T.}~\bibnamefont {Gog}}, \ and\
  \bibinfo {author} {\bibfnamefont {Y.-J.}\ \bibnamefont {Kim}},\ }\href
  {\doibase 10.1103/PhysRevB.83.075120} {\bibfield  {journal} {\bibinfo
  {journal} {Phys. Rev. B}\ }\textbf {\bibinfo {volume} {83}},\ \bibinfo
  {pages} {075120} (\bibinfo {year} {2011})}\BibitemShut {NoStop}%
\bibitem [{\citenamefont {Okada}\ and\ \citenamefont
  {Kotani}(2002)}]{Okada_2002_OKRIXSD4h}%
  \BibitemOpen
  \bibfield  {author} {\bibinfo {author} {\bibfnamefont {K.}~\bibnamefont
  {Okada}}\ and\ \bibinfo {author} {\bibfnamefont {A.}~\bibnamefont {Kotani}},\
  }\href {\doibase 10.1103/PhysRevB.65.144530} {\bibfield  {journal} {\bibinfo
  {journal} {Phys. Rev. B}\ }\textbf {\bibinfo {volume} {65}},\ \bibinfo
  {pages} {144530} (\bibinfo {year} {2002})}\BibitemShut {NoStop}%
\bibitem [{\citenamefont {Okada}\ and\ \citenamefont
  {Kotani}(2001{\natexlab{a}})}]{Okada_2001}%
  \BibitemOpen
  \bibfield  {author} {\bibinfo {author} {\bibfnamefont {K.}~\bibnamefont
  {Okada}}\ and\ \bibinfo {author} {\bibfnamefont {A.}~\bibnamefont {Kotani}},\
  }\href@noop {} {\bibfield  {journal} {\bibinfo  {journal} {J. Synchrotron
  Radiat.}\ } (\bibinfo {year} {2001}{\natexlab{a}})}\BibitemShut {NoStop}%
\bibitem [{\citenamefont {Okada}\ and\ \citenamefont
  {Kotani}(2001{\natexlab{b}})}]{Okada_2002}%
  \BibitemOpen
  \bibfield  {author} {\bibinfo {author} {\bibfnamefont {K.}~\bibnamefont
  {Okada}}\ and\ \bibinfo {author} {\bibfnamefont {A.}~\bibnamefont {Kotani}},\
  }\href {\doibase 10.1103/PhysRevB.63.045103} {\bibfield  {journal} {\bibinfo
  {journal} {Phys. Rev. B}\ }\textbf {\bibinfo {volume} {63}},\ \bibinfo
  {pages} {045103} (\bibinfo {year} {2001}{\natexlab{b}})}\BibitemShut
  {NoStop}%
\bibitem [{\citenamefont {Okada}\ and\ \citenamefont
  {Kotani}(2003)}]{Okada_2003}%
  \BibitemOpen
  \bibfield  {author} {\bibinfo {author} {\bibfnamefont {K.}~\bibnamefont
  {Okada}}\ and\ \bibinfo {author} {\bibfnamefont {A.}~\bibnamefont {Kotani}},\
  }\href {\doibase 10.1143/JPSJ.72.797} {\bibfield  {journal} {\bibinfo
  {journal} {J. Phys. Soc. Jpn.}\ }\textbf {\bibinfo {volume} {72}},\ \bibinfo
  {pages} {797} (\bibinfo {year} {2003})}\BibitemShut {NoStop}%
\bibitem [{\citenamefont {Okada}\ and\ \citenamefont
  {Kotani}(2006)}]{Okada_2006}%
  \BibitemOpen
  \bibfield  {author} {\bibinfo {author} {\bibfnamefont {K.}~\bibnamefont
  {Okada}}\ and\ \bibinfo {author} {\bibfnamefont {A.}~\bibnamefont {Kotani}},\
  }\href {\doibase 10.1143/JPSJ.75.044702} {\bibfield  {journal} {\bibinfo
  {journal} {J. Phys. Soc. Jpn.}\ }\textbf {\bibinfo {volume} {75}},\ \bibinfo
  {pages} {1} (\bibinfo {year} {2006})}\BibitemShut {NoStop}%
\bibitem [{\citenamefont {Okada}\ and\ \citenamefont
  {Kotani}(2007)}]{Okada_2007_OKCuL}%
  \BibitemOpen
  \bibfield  {author} {\bibinfo {author} {\bibfnamefont {K.}~\bibnamefont
  {Okada}}\ and\ \bibinfo {author} {\bibfnamefont {A.}~\bibnamefont {Kotani}},\
  }\href {https://doi.org/10.1143/JPSJ.76.123706} {\bibfield  {journal}
  {\bibinfo  {journal} {J. Phys. Soc. J}\ }\textbf {\bibinfo {volume} {76}},\
  \bibinfo {pages} {123706} (\bibinfo {year} {2007})}\BibitemShut {NoStop}%
\bibitem [{\citenamefont {Bisogni}\ \emph
  {et~al.}(2012{\natexlab{a}})\citenamefont {Bisogni}, \citenamefont
  {Simonelli}, \citenamefont {Ament}, \citenamefont {Forte}, \citenamefont
  {Moretti~Sala}, \citenamefont {Minola}, \citenamefont {Huotari},
  \citenamefont {van~den Brink}, \citenamefont {Ghiringhelli}, \citenamefont
  {Brookes},\ and\ \citenamefont {Braicovich}}]{Bisogni_2012_BimagnonStudyOKL}%
  \BibitemOpen
  \bibfield  {author} {\bibinfo {author} {\bibfnamefont {V.}~\bibnamefont
  {Bisogni}}, \bibinfo {author} {\bibfnamefont {L.}~\bibnamefont {Simonelli}},
  \bibinfo {author} {\bibfnamefont {L.~J.~P.}\ \bibnamefont {Ament}}, \bibinfo
  {author} {\bibfnamefont {F.}~\bibnamefont {Forte}}, \bibinfo {author}
  {\bibfnamefont {M.}~\bibnamefont {Moretti~Sala}}, \bibinfo {author}
  {\bibfnamefont {M.}~\bibnamefont {Minola}}, \bibinfo {author} {\bibfnamefont
  {S.}~\bibnamefont {Huotari}}, \bibinfo {author} {\bibfnamefont
  {J.}~\bibnamefont {van~den Brink}}, \bibinfo {author} {\bibfnamefont
  {G.}~\bibnamefont {Ghiringhelli}}, \bibinfo {author} {\bibfnamefont {N.~B.}\
  \bibnamefont {Brookes}}, \ and\ \bibinfo {author} {\bibfnamefont
  {L.}~\bibnamefont {Braicovich}},\ }\href {\doibase
  10.1103/PhysRevB.85.214527} {\bibfield  {journal} {\bibinfo  {journal} {Phys.
  Rev. B}\ }\textbf {\bibinfo {volume} {85}},\ \bibinfo {pages} {214527}
  (\bibinfo {year} {2012}{\natexlab{a}})}\BibitemShut {NoStop}%
\bibitem [{\citenamefont {Bisogni}\ \emph
  {et~al.}(2012{\natexlab{b}})\citenamefont {Bisogni}, \citenamefont {{Moretti
  Sala}}, \citenamefont {Bendounan}, \citenamefont {Brookes}, \citenamefont
  {Ghiringhelli},\ and\ \citenamefont
  {Braicovich}}]{Bisogni_2012_BimagnonStudyOKLII}%
  \BibitemOpen
  \bibfield  {author} {\bibinfo {author} {\bibfnamefont {V.}~\bibnamefont
  {Bisogni}}, \bibinfo {author} {\bibfnamefont {M.}~\bibnamefont {{Moretti
  Sala}}}, \bibinfo {author} {\bibfnamefont {A.}~\bibnamefont {Bendounan}},
  \bibinfo {author} {\bibfnamefont {N.~B.}\ \bibnamefont {Brookes}}, \bibinfo
  {author} {\bibfnamefont {G.}~\bibnamefont {Ghiringhelli}}, \ and\ \bibinfo
  {author} {\bibfnamefont {L.}~\bibnamefont {Braicovich}},\ }\href {\doibase
  10.1103/PhysRevB.85.214528} {\bibfield  {journal} {\bibinfo  {journal} {Phys.
  Rev. B}\ }\textbf {\bibinfo {volume} {85}},\ \bibinfo {pages} {214528}
  (\bibinfo {year} {2012}{\natexlab{b}})}\BibitemShut {NoStop}%
\bibitem [{\citenamefont {Monney}\ \emph {et~al.}(2013)\citenamefont {Monney},
  \citenamefont {Bisogni}, \citenamefont {Zhou}, \citenamefont {Kraus},
  \citenamefont {Strocov}, \citenamefont {Behr}, \citenamefont {M{\'{a}}lek},
  \citenamefont {Kuzian}, \citenamefont {Drechsler}, \citenamefont {Johnston},
  \citenamefont {Revcolevschi}, \citenamefont {B{\"{u}}chner}, \citenamefont
  {R{\o}nnow}, \citenamefont {{Van Den Brink}}, \citenamefont {Geck},\ and\
  \citenamefont {Schmitt}}]{Monney2013}%
  \BibitemOpen
  \bibfield  {author} {\bibinfo {author} {\bibfnamefont {C.}~\bibnamefont
  {Monney}}, \bibinfo {author} {\bibfnamefont {V.}~\bibnamefont {Bisogni}},
  \bibinfo {author} {\bibfnamefont {K.~J.}\ \bibnamefont {Zhou}}, \bibinfo
  {author} {\bibfnamefont {R.}~\bibnamefont {Kraus}}, \bibinfo {author}
  {\bibfnamefont {V.~N.}\ \bibnamefont {Strocov}}, \bibinfo {author}
  {\bibfnamefont {G.}~\bibnamefont {Behr}}, \bibinfo {author} {\bibfnamefont
  {J.}~\bibnamefont {M{\'{a}}lek}}, \bibinfo {author} {\bibfnamefont
  {R.}~\bibnamefont {Kuzian}}, \bibinfo {author} {\bibfnamefont {S.~L.}\
  \bibnamefont {Drechsler}}, \bibinfo {author} {\bibfnamefont {S.}~\bibnamefont
  {Johnston}}, \bibinfo {author} {\bibfnamefont {A.}~\bibnamefont
  {Revcolevschi}}, \bibinfo {author} {\bibfnamefont {B.}~\bibnamefont
  {B{\"{u}}chner}}, \bibinfo {author} {\bibfnamefont {H.~M.}\ \bibnamefont
  {R{\o}nnow}}, \bibinfo {author} {\bibfnamefont {J.}~\bibnamefont {{Van Den
  Brink}}}, \bibinfo {author} {\bibfnamefont {J.}~\bibnamefont {Geck}}, \ and\
  \bibinfo {author} {\bibfnamefont {T.}~\bibnamefont {Schmitt}},\ }\href
  {\doibase 10.1103/PhysRevLett.110.087403} {\bibfield  {journal} {\bibinfo
  {journal} {Phys. Rev. Lett.}\ }\textbf {\bibinfo {volume} {110}},\ \bibinfo
  {pages} {1} (\bibinfo {year} {2013})}\BibitemShut {NoStop}%
\bibitem [{\citenamefont {Ishii}\ \emph {et~al.}(2017)\citenamefont {Ishii},
  \citenamefont {Tohyama}, \citenamefont {Asano}, \citenamefont {Sato},
  \citenamefont {Fujita}, \citenamefont {Wakimoto}, \citenamefont {Tustsui},
  \citenamefont {Sota}, \citenamefont {Miyawaki}, \citenamefont {Niwa},
  \citenamefont {Harada}, \citenamefont {Pelliciari}, \citenamefont {Huang},
  \citenamefont {Schmitt}, \citenamefont {Yamamoto},\ and\ \citenamefont
  {Mizuki}}]{Ishii_2017_OKexp}%
  \BibitemOpen
  \bibfield  {author} {\bibinfo {author} {\bibfnamefont {K.}~\bibnamefont
  {Ishii}}, \bibinfo {author} {\bibfnamefont {T.}~\bibnamefont {Tohyama}},
  \bibinfo {author} {\bibfnamefont {S.}~\bibnamefont {Asano}}, \bibinfo
  {author} {\bibfnamefont {K.}~\bibnamefont {Sato}}, \bibinfo {author}
  {\bibfnamefont {M.}~\bibnamefont {Fujita}}, \bibinfo {author} {\bibfnamefont
  {S.}~\bibnamefont {Wakimoto}}, \bibinfo {author} {\bibfnamefont
  {K.}~\bibnamefont {Tustsui}}, \bibinfo {author} {\bibfnamefont
  {S.}~\bibnamefont {Sota}}, \bibinfo {author} {\bibfnamefont {J.}~\bibnamefont
  {Miyawaki}}, \bibinfo {author} {\bibfnamefont {H.}~\bibnamefont {Niwa}},
  \bibinfo {author} {\bibfnamefont {Y.}~\bibnamefont {Harada}}, \bibinfo
  {author} {\bibfnamefont {J.}~\bibnamefont {Pelliciari}}, \bibinfo {author}
  {\bibfnamefont {Y.}~\bibnamefont {Huang}}, \bibinfo {author} {\bibfnamefont
  {T.}~\bibnamefont {Schmitt}}, \bibinfo {author} {\bibfnamefont
  {Y.}~\bibnamefont {Yamamoto}}, \ and\ \bibinfo {author} {\bibfnamefont
  {J.}~\bibnamefont {Mizuki}},\ }\href {\doibase 10.1103/PhysRevB.96.115148}
  {\bibfield  {journal} {\bibinfo  {journal} {Phys. Rev. B}\ }\textbf {\bibinfo
  {volume} {96}},\ \bibinfo {pages} {115148} (\bibinfo {year}
  {2017})}\BibitemShut {NoStop}%
\bibitem [{\citenamefont {Schlappa}\ \emph {et~al.}(2018)\citenamefont
  {Schlappa}, \citenamefont {Kumar}, \citenamefont {Zhou}, \citenamefont
  {Singh}, \citenamefont {Mourigal}, \citenamefont {Strocov}, \citenamefont
  {Revcolevschi}, \citenamefont {Patthey}, \citenamefont {R{\o}nnow},
  \citenamefont {Johnston},\ and\ \citenamefont
  {Schmitt}}]{Schlappa_2018_OK_Twospinon}%
  \BibitemOpen
  \bibfield  {author} {\bibinfo {author} {\bibfnamefont {J.}~\bibnamefont
  {Schlappa}}, \bibinfo {author} {\bibfnamefont {U.}~\bibnamefont {Kumar}},
  \bibinfo {author} {\bibfnamefont {K.~J.}\ \bibnamefont {Zhou}}, \bibinfo
  {author} {\bibfnamefont {S.}~\bibnamefont {Singh}}, \bibinfo {author}
  {\bibfnamefont {M.}~\bibnamefont {Mourigal}}, \bibinfo {author}
  {\bibfnamefont {V.~N.}\ \bibnamefont {Strocov}}, \bibinfo {author}
  {\bibfnamefont {A.}~\bibnamefont {Revcolevschi}}, \bibinfo {author}
  {\bibfnamefont {L.}~\bibnamefont {Patthey}}, \bibinfo {author} {\bibfnamefont
  {H.~M.}\ \bibnamefont {R{\o}nnow}}, \bibinfo {author} {\bibfnamefont
  {S.}~\bibnamefont {Johnston}}, \ and\ \bibinfo {author} {\bibfnamefont
  {T.}~\bibnamefont {Schmitt}},\ }\href {\doibase 10.1038/s41467-018-07838-y}
  {\bibfield  {journal} {\bibinfo  {journal} {Nat. Commun.}\ }\textbf {\bibinfo
  {volume} {9}},\ \bibinfo {pages} {5394} (\bibinfo {year} {2018})}\BibitemShut
  {NoStop}%
\bibitem [{\citenamefont {Yamagami}\ \emph {et~al.}(2020)\citenamefont
  {Yamagami}, \citenamefont {Ishii}, \citenamefont {Hirata}, \citenamefont
  {Ikeda}, \citenamefont {Miyawaki}, \citenamefont {Harada}, \citenamefont
  {Miyazaki}, \citenamefont {Asano}, \citenamefont {Fujita},\ and\
  \citenamefont {Wadati}}]{Yamagami_2020_OKexp}%
  \BibitemOpen
  \bibfield  {author} {\bibinfo {author} {\bibfnamefont {K.}~\bibnamefont
  {Yamagami}}, \bibinfo {author} {\bibfnamefont {K.}~\bibnamefont {Ishii}},
  \bibinfo {author} {\bibfnamefont {Y.}~\bibnamefont {Hirata}}, \bibinfo
  {author} {\bibfnamefont {K.}~\bibnamefont {Ikeda}}, \bibinfo {author}
  {\bibfnamefont {J.}~\bibnamefont {Miyawaki}}, \bibinfo {author}
  {\bibfnamefont {Y.}~\bibnamefont {Harada}}, \bibinfo {author} {\bibfnamefont
  {M.}~\bibnamefont {Miyazaki}}, \bibinfo {author} {\bibfnamefont
  {S.}~\bibnamefont {Asano}}, \bibinfo {author} {\bibfnamefont
  {M.}~\bibnamefont {Fujita}}, \ and\ \bibinfo {author} {\bibfnamefont
  {H.}~\bibnamefont {Wadati}},\ }\href {\doibase 10.1103/PhysRevB.102.165145}
  {\bibfield  {journal} {\bibinfo  {journal} {Phys. Rev. B}\ }\textbf {\bibinfo
  {volume} {102}},\ \bibinfo {pages} {165145} (\bibinfo {year}
  {2020})}\BibitemShut {NoStop}%
\bibitem [{\citenamefont {Paris}\ \emph {et~al.}(2021)\citenamefont {Paris},
  \citenamefont {Nicholson}, \citenamefont {Johnston}, \citenamefont {Tseng},
  \citenamefont {Rumo}, \citenamefont {Coslovich}, \citenamefont {Zohar},
  \citenamefont {Lin}, \citenamefont {Strocov}, \citenamefont {Saint-Martin},
  \citenamefont {Revcolevschi}, \citenamefont {Kemper}, \citenamefont
  {Schlotter}, \citenamefont {Dakovski}, \citenamefont {Monney},\ and\
  \citenamefont {Schmitt}}]{Paris_2021_OK_CGO}%
  \BibitemOpen
  \bibfield  {author} {\bibinfo {author} {\bibfnamefont {E.}~\bibnamefont
  {Paris}}, \bibinfo {author} {\bibfnamefont {C.~W.}\ \bibnamefont
  {Nicholson}}, \bibinfo {author} {\bibfnamefont {S.}~\bibnamefont {Johnston}},
  \bibinfo {author} {\bibfnamefont {Y.}~\bibnamefont {Tseng}}, \bibinfo
  {author} {\bibfnamefont {M.}~\bibnamefont {Rumo}}, \bibinfo {author}
  {\bibfnamefont {G.}~\bibnamefont {Coslovich}}, \bibinfo {author}
  {\bibfnamefont {S.}~\bibnamefont {Zohar}}, \bibinfo {author} {\bibfnamefont
  {M.~F.}\ \bibnamefont {Lin}}, \bibinfo {author} {\bibfnamefont {V.~N.}\
  \bibnamefont {Strocov}}, \bibinfo {author} {\bibfnamefont {R.}~\bibnamefont
  {Saint-Martin}}, \bibinfo {author} {\bibfnamefont {A.}~\bibnamefont
  {Revcolevschi}}, \bibinfo {author} {\bibfnamefont {A.}~\bibnamefont
  {Kemper}}, \bibinfo {author} {\bibfnamefont {W.}~\bibnamefont {Schlotter}},
  \bibinfo {author} {\bibfnamefont {G.~L.}\ \bibnamefont {Dakovski}}, \bibinfo
  {author} {\bibfnamefont {C.}~\bibnamefont {Monney}}, \ and\ \bibinfo {author}
  {\bibfnamefont {T.}~\bibnamefont {Schmitt}},\ }\href {\doibase
  10.1038/s41535-021-00350-5} {\bibfield  {journal} {\bibinfo  {journal} {npj
  Quantum Materials}\ }\textbf {\bibinfo {volume} {6}},\ \bibinfo {pages} {51}
  (\bibinfo {year} {2021})}\BibitemShut {NoStop}%
\bibitem [{\citenamefont {Shen}\ \emph {et~al.}(2022)\citenamefont {Shen},
  \citenamefont {Sears}, \citenamefont {Fabbris}, \citenamefont {Li},
  \citenamefont {Pelliciari}, \citenamefont {Jarrige}, \citenamefont {He},
  \citenamefont {Bo\ifmmode \check{z}\else \v{z}\fi{}ovi\ifmmode~\acute{c}\else
  \'{c}\fi{}}, \citenamefont {Mitrano}, \citenamefont {Zhang}, \citenamefont
  {Mitchell}, \citenamefont {Botana}, \citenamefont {Bisogni}, \citenamefont
  {Norman}, \citenamefont {Johnston},\ and\ \citenamefont
  {Dean}}]{Shen_2022_OKRIXS_LNO}%
  \BibitemOpen
  \bibfield  {author} {\bibinfo {author} {\bibfnamefont {Y.}~\bibnamefont
  {Shen}}, \bibinfo {author} {\bibfnamefont {J.}~\bibnamefont {Sears}},
  \bibinfo {author} {\bibfnamefont {G.}~\bibnamefont {Fabbris}}, \bibinfo
  {author} {\bibfnamefont {J.}~\bibnamefont {Li}}, \bibinfo {author}
  {\bibfnamefont {J.}~\bibnamefont {Pelliciari}}, \bibinfo {author}
  {\bibfnamefont {I.}~\bibnamefont {Jarrige}}, \bibinfo {author} {\bibfnamefont
  {X.}~\bibnamefont {He}}, \bibinfo {author} {\bibfnamefont {I.}~\bibnamefont
  {Bo\ifmmode \check{z}\else \v{z}\fi{}ovi\ifmmode~\acute{c}\else \'{c}\fi{}}},
  \bibinfo {author} {\bibfnamefont {M.}~\bibnamefont {Mitrano}}, \bibinfo
  {author} {\bibfnamefont {J.}~\bibnamefont {Zhang}}, \bibinfo {author}
  {\bibfnamefont {J.~F.}\ \bibnamefont {Mitchell}}, \bibinfo {author}
  {\bibfnamefont {A.~S.}\ \bibnamefont {Botana}}, \bibinfo {author}
  {\bibfnamefont {V.}~\bibnamefont {Bisogni}}, \bibinfo {author} {\bibfnamefont
  {M.~R.}\ \bibnamefont {Norman}}, \bibinfo {author} {\bibfnamefont
  {S.}~\bibnamefont {Johnston}}, \ and\ \bibinfo {author} {\bibfnamefont
  {M.~P.~M.}\ \bibnamefont {Dean}},\ }\href {\doibase
  10.1103/PhysRevX.12.011055} {\bibfield  {journal} {\bibinfo  {journal} {Phys.
  Rev. X}\ }\textbf {\bibinfo {volume} {12}},\ \bibinfo {pages} {011055}
  (\bibinfo {year} {2022})}\BibitemShut {NoStop}%
\bibitem [{\citenamefont {Tsutsui}\ \emph {et~al.}(1999)\citenamefont
  {Tsutsui}, \citenamefont {Tohyama},\ and\ \citenamefont
  {Maekawa}}]{TsutsuiTohyama_1999_2dCuK}%
  \BibitemOpen
  \bibfield  {author} {\bibinfo {author} {\bibfnamefont {K.}~\bibnamefont
  {Tsutsui}}, \bibinfo {author} {\bibfnamefont {T.}~\bibnamefont {Tohyama}}, \
  and\ \bibinfo {author} {\bibfnamefont {S.}~\bibnamefont {Maekawa}},\ }\href
  {\doibase 10.1103/PhysRevLett.83.3705} {\bibfield  {journal} {\bibinfo
  {journal} {Phys. Rev. Lett.}\ }\textbf {\bibinfo {volume} {83}},\ \bibinfo
  {pages} {3705} (\bibinfo {year} {1999})}\BibitemShut {NoStop}%
\bibitem [{\citenamefont {Tsutsui}\ \emph {et~al.}(2000)\citenamefont
  {Tsutsui}, \citenamefont {Tohyama},\ and\ \citenamefont
  {Maekawa}}]{TsutsuiTohyama_2000_1dCuK}%
  \BibitemOpen
  \bibfield  {author} {\bibinfo {author} {\bibfnamefont {K.}~\bibnamefont
  {Tsutsui}}, \bibinfo {author} {\bibfnamefont {T.}~\bibnamefont {Tohyama}}, \
  and\ \bibinfo {author} {\bibfnamefont {S.}~\bibnamefont {Maekawa}},\ }\href
  {\doibase 10.1103/PhysRevB.61.7180} {\bibfield  {journal} {\bibinfo
  {journal} {Phys. Rev. B}\ }\textbf {\bibinfo {volume} {61}},\ \bibinfo
  {pages} {7180} (\bibinfo {year} {2000})}\BibitemShut {NoStop}%
\bibitem [{\citenamefont {Chen}\ \emph {et~al.}(2010)\citenamefont {Chen},
  \citenamefont {Moritz}, \citenamefont {Vernay}, \citenamefont {Hancock},
  \citenamefont {Johnston}, \citenamefont {Jia}, \citenamefont
  {Chabot-Couture}, \citenamefont {Greven}, \citenamefont {Elfimov},
  \citenamefont {Sawatzky},\ and\ \citenamefont
  {Devereaux}}]{Chen_2010_CuK_RIXS_Theory}%
  \BibitemOpen
  \bibfield  {author} {\bibinfo {author} {\bibfnamefont {C.-C.}\ \bibnamefont
  {Chen}}, \bibinfo {author} {\bibfnamefont {B.}~\bibnamefont {Moritz}},
  \bibinfo {author} {\bibfnamefont {F.}~\bibnamefont {Vernay}}, \bibinfo
  {author} {\bibfnamefont {J.~N.}\ \bibnamefont {Hancock}}, \bibinfo {author}
  {\bibfnamefont {S.}~\bibnamefont {Johnston}}, \bibinfo {author}
  {\bibfnamefont {C.~J.}\ \bibnamefont {Jia}}, \bibinfo {author} {\bibfnamefont
  {G.}~\bibnamefont {Chabot-Couture}}, \bibinfo {author} {\bibfnamefont
  {M.}~\bibnamefont {Greven}}, \bibinfo {author} {\bibfnamefont
  {I.}~\bibnamefont {Elfimov}}, \bibinfo {author} {\bibfnamefont {G.~A.}\
  \bibnamefont {Sawatzky}}, \ and\ \bibinfo {author} {\bibfnamefont {T.~P.}\
  \bibnamefont {Devereaux}},\ }\href {\doibase 10.1103/PhysRevLett.105.177401}
  {\bibfield  {journal} {\bibinfo  {journal} {Phys. Rev. Lett.}\ }\textbf
  {\bibinfo {volume} {105}},\ \bibinfo {pages} {177401} (\bibinfo {year}
  {2010})}\BibitemShut {NoStop}%
\bibitem [{\citenamefont {Kourtis}\ \emph {et~al.}(2011)\citenamefont
  {Kourtis}, \citenamefont {van~den Brink},\ and\ \citenamefont
  {Daghofer}}]{Kourtis_2012_1dMott_CuK}%
  \BibitemOpen
  \bibfield  {author} {\bibinfo {author} {\bibfnamefont {S.}~\bibnamefont
  {Kourtis}}, \bibinfo {author} {\bibfnamefont {J.}~\bibnamefont {van~den
  Brink}}, \ and\ \bibinfo {author} {\bibfnamefont {M.}~\bibnamefont
  {Daghofer}},\ }\href {http://arxiv.org/abs/1111.6028
  http://dx.doi.org/10.1103/PhysRevB.85.064423} {\bibfield  {journal} {\bibinfo
   {journal} {Phys. Rev. B}\ }\textbf {\bibinfo {volume} {85}} (\bibinfo {year}
  {2011})}\BibitemShut {NoStop}%
\bibitem [{\citenamefont {Nomura}\ and\ \citenamefont
  {Igarashi}(2005)}]{Igarashi_2005_K-edge1}%
  \BibitemOpen
  \bibfield  {author} {\bibinfo {author} {\bibfnamefont {T.}~\bibnamefont
  {Nomura}}\ and\ \bibinfo {author} {\bibfnamefont {J.}~\bibnamefont
  {Igarashi}},\ }\href {\doibase 10.1103/PhysRevB.71.035110} {\bibfield
  {journal} {\bibinfo  {journal} {Phys. Rev. B}\ }\textbf {\bibinfo {volume}
  {71}},\ \bibinfo {pages} {035110} (\bibinfo {year} {2005})}\BibitemShut
  {NoStop}%
\bibitem [{\citenamefont {Igarashi}\ \emph {et~al.}(2006)\citenamefont
  {Igarashi}, \citenamefont {Nomura},\ and\ \citenamefont
  {Takahashi}}]{Igarashi_2006_K-edge2}%
  \BibitemOpen
  \bibfield  {author} {\bibinfo {author} {\bibfnamefont {J.}~\bibnamefont
  {Igarashi}}, \bibinfo {author} {\bibfnamefont {T.}~\bibnamefont {Nomura}}, \
  and\ \bibinfo {author} {\bibfnamefont {M.}~\bibnamefont {Takahashi}},\ }\href
  {\doibase 10.1103/PhysRevB.74.245122} {\bibfield  {journal} {\bibinfo
  {journal} {Phys. Rev. B}\ }\textbf {\bibinfo {volume} {74}},\ \bibinfo
  {pages} {245122} (\bibinfo {year} {2006})}\BibitemShut {NoStop}%
\bibitem [{\citenamefont {Igarashi}\ and\ \citenamefont
  {Nagao}(2013)}]{Igarashi_2013_L-edge}%
  \BibitemOpen
  \bibfield  {author} {\bibinfo {author} {\bibfnamefont {J.}~\bibnamefont
  {Igarashi}}\ and\ \bibinfo {author} {\bibfnamefont {T.}~\bibnamefont
  {Nagao}},\ }\href {\doibase 10.1103/PhysRevB.88.014407} {\bibfield  {journal}
  {\bibinfo  {journal} {Phys. Rev. B}\ }\textbf {\bibinfo {volume} {88}},\
  \bibinfo {pages} {014407} (\bibinfo {year} {2013})}\BibitemShut {NoStop}%
\bibitem [{\citenamefont {Yin}\ \emph {et~al.}(2008)\citenamefont {Yin},
  \citenamefont {Gordienko}, \citenamefont {Wan},\ and\ \citenamefont
  {Savrasov}}]{Yin2008_ZRS_DMFT}%
  \BibitemOpen
  \bibfield  {author} {\bibinfo {author} {\bibfnamefont {Q.}~\bibnamefont
  {Yin}}, \bibinfo {author} {\bibfnamefont {A.}~\bibnamefont {Gordienko}},
  \bibinfo {author} {\bibfnamefont {X.}~\bibnamefont {Wan}}, \ and\ \bibinfo
  {author} {\bibfnamefont {S.~Y.}\ \bibnamefont {Savrasov}},\ }\href {\doibase
  10.1103/PhysRevLett.100.066406} {\bibfield  {journal} {\bibinfo  {journal}
  {Phys. Rev. Lett.}\ }\textbf {\bibinfo {volume} {100}},\ \bibinfo {pages}
  {066406} (\bibinfo {year} {2008})}\BibitemShut {NoStop}%
\bibitem [{\citenamefont {Weber}\ \emph {et~al.}(2010)\citenamefont {Weber},
  \citenamefont {Haule},\ and\ \citenamefont {Kotliar}}]{Weber2010}%
  \BibitemOpen
  \bibfield  {author} {\bibinfo {author} {\bibfnamefont {C.}~\bibnamefont
  {Weber}}, \bibinfo {author} {\bibfnamefont {K.}~\bibnamefont {Haule}}, \ and\
  \bibinfo {author} {\bibfnamefont {G.}~\bibnamefont {Kotliar}},\ }\href
  {\doibase 10.1103/PhysRevB.82.125107} {\bibfield  {journal} {\bibinfo
  {journal} {Phys. Rev. B}\ }\textbf {\bibinfo {volume} {82}},\ \bibinfo
  {pages} {125107} (\bibinfo {year} {2010})}\BibitemShut {NoStop}%
\bibitem [{\citenamefont {Cui}\ \emph {et~al.}(2020)\citenamefont {Cui},
  \citenamefont {Sun}, \citenamefont {Ray}, \citenamefont {Zheng},
  \citenamefont {Sun},\ and\ \citenamefont {Chan}}]{Cui2020_dp}%
  \BibitemOpen
  \bibfield  {author} {\bibinfo {author} {\bibfnamefont {Z.~H.}\ \bibnamefont
  {Cui}}, \bibinfo {author} {\bibfnamefont {C.}~\bibnamefont {Sun}}, \bibinfo
  {author} {\bibfnamefont {U.}~\bibnamefont {Ray}}, \bibinfo {author}
  {\bibfnamefont {B.~X.}\ \bibnamefont {Zheng}}, \bibinfo {author}
  {\bibfnamefont {Q.}~\bibnamefont {Sun}}, \ and\ \bibinfo {author}
  {\bibfnamefont {G.~K.~L.}\ \bibnamefont {Chan}},\ }\href {\doibase
  10.1103/PhysRevResearch.2.043259} {\bibfield  {journal} {\bibinfo  {journal}
  {Phys. Rev. Res.}\ }\textbf {\bibinfo {volume} {2}},\ \bibinfo {pages}
  {043259} (\bibinfo {year} {2020})}\BibitemShut {NoStop}%
\bibitem [{\citenamefont {S{\'{e}}n{\'{e}}chal}\ \emph
  {et~al.}(2000)\citenamefont {S{\'{e}}n{\'{e}}chal}, \citenamefont {Perez},\
  and\ \citenamefont {Pioro-Ladri{\`{e}}re}}]{Senechal_2000}%
  \BibitemOpen
  \bibfield  {author} {\bibinfo {author} {\bibfnamefont {D.}~\bibnamefont
  {S{\'{e}}n{\'{e}}chal}}, \bibinfo {author} {\bibfnamefont {D.}~\bibnamefont
  {Perez}}, \ and\ \bibinfo {author} {\bibfnamefont {M.}~\bibnamefont
  {Pioro-Ladri{\`{e}}re}},\ }\href {\doibase 10.1103/PhysRevLett.84.522}
  {\bibfield  {journal} {\bibinfo  {journal} {Phys. Rev. Lett.}\ }\textbf
  {\bibinfo {volume} {84}},\ \bibinfo {pages} {522} (\bibinfo {year}
  {2000})}\BibitemShut {NoStop}%
\bibitem [{\citenamefont {S{\'{e}}n{\'{e}}chal}\ \emph
  {et~al.}(2002)\citenamefont {S{\'{e}}n{\'{e}}chal}, \citenamefont
  {P{\'{e}}rez},\ and\ \citenamefont {Plouffe}}]{Senechal_2002_CPTHubbard}%
  \BibitemOpen
  \bibfield  {author} {\bibinfo {author} {\bibfnamefont {D.}~\bibnamefont
  {S{\'{e}}n{\'{e}}chal}}, \bibinfo {author} {\bibfnamefont {D.}~\bibnamefont
  {P{\'{e}}rez}}, \ and\ \bibinfo {author} {\bibfnamefont {D.}~\bibnamefont
  {Plouffe}},\ }\href {\doibase 10.1103/PhysRevB.66.075129} {\bibfield
  {journal} {\bibinfo  {journal} {Phys. Rev. B}\ }\textbf {\bibinfo {volume}
  {66}},\ \bibinfo {pages} {751291} (\bibinfo {year} {2002})}\BibitemShut
  {NoStop}%
\end{thebibliography}%
\end{document}